\documentclass[fleqn,usenatbib]{mnras}

\usepackage{newtxtext,newtxmath}

\usepackage[T1]{fontenc}

\DeclareRobustCommand{\VAN}[3]{#2}
\let\VANthebibliography\thebibliography
\def\thebibliography{\DeclareRobustCommand{\VAN}[3]{##3}\VANthebibliography}

\usepackage{graphicx}  

\newcommand\asca{\textit{ASCA}}
\newcommand\beppo{\textit{BeppoSAX}}
\newcommand\rxte{\textit{RXTE}}
\newcommand\chandra{\textit{Chandra}}
\newcommand\xmm{\textit{XMM-Newton}}
\newcommand\nustar{\textit{NuSTAR}}
\newcommand\suzaku{\textit{Suzaku}}
\newcommand\swift{\textit{Swift}}
\newcommand\nh{N_{\rm H}}
\newcommand{\nodata}{~$\cdots$~}
\defcitealias{2013Sci...341..981W}{W13}

\title[X-ray variation in M81*]{The X-ray variation of M81* resolved by \chandra\ and \nustar}

\author[Niu et al.]{
  Shu Niu,$^{1,2,}$\thanks{E-mail: niushu@pmo.ac.cn (S.N.), fgxie@shao.ac.cn (F.G.X.), ji@pmo.ac.cn (L.J.)}
  Fu-Guo Xie,$^{1,\star}$
  Q. Daniel Wang,$^{3}$
  Li Ji,$^{4,5,\star}$
  Feng Yuan,$^{1,2}$
  and Min Long$^{6}$
  \\
  $^{1}$Key Laboratory for Research in Galaxies and Cosmology, Shanghai Astronomical Observatory, Chinese Academy of Sciences, \\
  Shanghai 200030, People's Republic of China\\
  $^{2}$University of Chinese Academy of Sciences, 19A Yuquan Road, Beijing 100049, People's Republic of China\\
  $^{3}$University of Massachusetts Amherst, Amherst, MA 01003, USA\\
  $^{4}$Purple Mountain Observatory, Chinese Academy of Sciences, Nanjing 210023, People's Republic of China\\
  $^{5}$Key Laboratory of Dark Matter and Space Astronomy, Chinese Academy of Sciences, People's Republic of China\\
  $^{6}$Department of Computer Science, Boise State University, Boise, ID 83725, USA
}

\date{Accepted XXX. Received YYY; in original form ZZZ}

\pubyear{2023}

\begin{document}
  \label{firstpage}
  \pagerange{\pageref{firstpage}--\pageref{lastpage}}
  \maketitle
  
  \begin{abstract}
    Despite advances in our understanding of low luminosity active galactic nuclei (LLAGNs), the fundamental details about the mechanisms of radiation and flare/outburst in hot accretion flow are still largely missing. We have systematically analyzed the archival \chandra\ and \nustar\ X-ray data of the nearby LLAGN M81*, whose $L_{\rm bol}\sim 10^{-5} L_{\rm Edd}$. 
    Through a detailed study of X-ray light curve and spectral properties, we find that the X-ray continuum emission of the power-law shape more likely originates from inverse Compton scattering  within the hot accretion flow. In contrast to Sgr A*, flares are rare in M81*. Low-amplitude variation can only be observed in soft X-ray band (amplitude usually $\lesssim 2$). Several simple models are tested, including sinusoidal-like and quasi-periodical. Based on a comparison of the dramatic differences of flare properties among Sgr A*, M31* and M81*, we find that, when the differences in both the accretion rate and the black hole mass are considered, the flares in LLAGNs can be understood universally in a magneto-hydrodynamical model.
  \end{abstract}

  
  
  
  \section{Introduction} \label{sec:intro}
  
  Consensus has been reached that, low-luminosity (LL) active galactic nuclei (AGNs), whose bolometric luminosity $L_{\rm bol}\lesssim(1-2)\%\,L_{\rm Edd}$,\footnote{The Eddington luminosity for accretion onto a black hole (BH) with mass $M_{\rm BH}$ is $L_{\rm Edd}\approx1.3\times10^{46}\, {\rm erg\,s^{-1}} (M_{\rm BH}/10^8 M_{\sun})$. Note that $L_{\rm Edd}\propto M_{\rm BH}$.} are distinctive to their bright/luminous cousins (e.g, \citealt{2008ARA&A..46..475H, 2014ARA&A..52..589H}), i.e., LLAGNs lack the big blue bump in optical/UV and the relativistic Fe line in X-rays. The reflection component in X-rays is also weak. Theoretically, luminous AGNs are believed to be powered by a cold thin accretion disk \citep{1973A&A....24..337S}, while LLAGNs are powered by a hot accretion flow (e.g., advection-dominated accretion flow [ADAF],  \citealt{1994ApJ...428L..13N}, for reviews, see \citealt{2008ARA&A..46..475H, 2014ARA&A..52..529Y}). Typically relativistic jets are also observed in LLAGNs. The significant change in accretion mode among LLAGNs and bright AGNs also leads to dramatic changes in AGN feedback (e.g., \citealt{2012ARA&A..50..455F, 2012NJPh...14e5023M, 2014ARA&A..52..589H}), where jet, wind and radiation from LLAGNs will have significant impacts on the gas dynamics (thus star formation) of the host galaxy.
  
  However, it remains unclear how the cold to hot (and reverse) accretion state transition actually occur (e.g., see \citealt{2014ARA&A..52..529Y} for a summary of various theoretical models). Accretion in some BH systems may be partly cold and partly hot. Moreover, in the theory of hot accretion flow, flares are expected to be generated by magnetic reconnection events (e.g. \citealt{2009MNRAS.395.2183Y, 2015ApJ...810...19L}, see Sec.\ \ref{sec:flare_model} below). Although the flares are indeed observed in the ultra low-luminosity system Sgr A* (Galactic center, e.g., \citealt{2009ApJ...702..178Y, 2013ApJ...774...42N}), their existence in other more brighter LLAGNs remains unclear yet. What drives such difference has not been systematically examined yet.
  
  These motivate us to investigate M81*, a nearby LLAGN (\citealt*{1981ApJ...245..845P, 1988ApJ...324..134F}) located at a distance $d=3.61^{+0.21}_{-0.19}$ Mpc \citep{2016AJ....152...50T}. M81* is also classified as a LINER \citep[low-ionization nuclear emission-line regions,][]{1980A&A....87..152H}. The bolometric luminosity of M81* is $L_{\rm bol}\approx 1.1\times10^{-5} L_{\rm Edd}$ \citep{1996ApJ...462..183H,2014MNRAS.438.2804N}, where the BH mass is $M_{\rm BH} = 7\times10^7 M_{\sun}$ \citep{2003AJ....125.1226D}. Thanks to its proximity in distance, several multi-wavelength monitoring campaigns have been carried out \citep[e.g.,][]{2008ApJ...681..905M, 2010ApJ...720.1033M}. A clear stationary core and a one-sided jet structure is observed \citep{2000ApJ...532..895B}. Moreover, a broad double-peaked H$\alpha$ feature is observed in optical, which suggests that outside of a truncation radius $R_{\rm tr}\sim 300\ R_s$ there exists a cold thin disk (\citealt{2007ApJ...671..118D}, see also \citealt{1999ApJ...525L..89Q,2009RAA.....9..401X}). Here $R_s = 2GM_{\rm BH}/c^2$ is the Schwarzschild radius of BH. 
  Considering the anti-correlation between $R_{\rm tr}$ and $L_{\rm bol}/L_{\rm Edd}$ in LLAGNs and BH binaries \citep[e.g.,][]{2004ApJ...612..724Y}, detailed studies of M81* will help to further our understanding in accretion theory, e.g., the formation of a cold thin disk.
  
  In this work, we focus on the X-ray properties of M81*. With a hydrogen column density $\nh\approx4.3-6.3\times10^{20}\,{\rm cm}^{-2}$ \citep{2018MNRAS.476.5698Y}, the X-ray continuum of M81* does not suffer much absorption.\footnote{Note that the Milky Way column density in the direction of M81* is $\nh ({\rm MW}) = 4.2\times10^{20}\,{\rm cm}^{-2}$ \citep{1990ARA&A..28..215D}.}
  The X-ray emission below $\sim$50-100 keV can be well described by a simple power-law $F_\epsilon\propto \epsilon^{1-\Gamma}$, where $\epsilon=h\nu$ is the photon energy. The photon index is found to be $\Gamma\sim1.7-1.8$ (\asca: \citealt{1996PASJ...48..237I}; \beppo: \citealt{2000A&A...353..447P}; \xmm: \citealt{2004A&A...422...77P}; \chandra: \citealt{2007ApJ...669..830Y}; \nustar\ and \suzaku: \citealt{2018MNRAS.476.5698Y}). However, the origin of this continuum power-law X-ray emission, either from the relativistic jet or from the hot accretion flow, remains inclusive 
  \citep[e.g.,][]{2008ApJ...681..905M, 2010ApJ...720.1033M, 2014MNRAS.438.2804N,2017ApJ...836..104X}. The hard X-rays by \nustar\ also show a cut-off at the energy $E_{\rm c}~ \sim$ 250 keV \citep{2018MNRAS.476.5698Y}.
  The X-ray emission in high spatial resolution \chandra\ observation also found to have a spatially diffuse component \citep{2003ApJS..144..213S}, similar to that in Sgr A*. However, the origin of these diffuse emission remains unclear.
  
  Fe K$\alpha$, Fe {\sc xxv} and Fe {\sc xxvi} emissions are also detected in X-rays. 
  Among them a receding velocity $\sim 2000-3000~{\rm km~s^{-1}}$ is detected in Fe {\sc xxvi} \citep{2004A&A...422...77P, 2007ApJ...669..830Y}. Recently \citet{2021NatAs...5..928S} argue that the Fe {\sc xxvi} lines are produced by hot outflows generated from hot accretion flow, i.e. meets a long theoretical expectation \citep[][and references therein]{2012ApJ...761..130Y,2015ApJ...804..101Y}. The Fe K$\alpha$ line  of neutral or weakly ionized iron, on the other hand, may originate from reflection of a truncated accretion disk or a torus of cold gas that are irradiated by the central LLAGN \citep{2018MNRAS.476.5698Y}.
  \citet{2018MNRAS.476.5698Y} carried out simultaneous observations in both soft and hard X-rays, where the Compton reflection hump is not detected (upper limit of $R<0.1$, and consistent with $R=0$). This is consistent with the observation that the cold thin disk is truncated at $\sim 300 R_s$.
  
  In this work, we present an analysis of archival broadband X-ray data by \chandra\ and \nustar. 
  The main motivation of this work is to investigate the accretion physics (i.e., origin of the X-ray emission, light curve, and possibly-detected flare) in M81*, with a comparison to those in a similar but much fainter system Sgr A* ($L_{\rm bol}\approx 1-2\times10^{-9} L_{\rm Edd}$, e.g, \citealt{2003ApJ...591..891B,2019MNRAS.483.5614M,2023ApJ...942...20X}).
  This work is organized as follows. Section \ref{sec:obs} illustrates the whole dataset and data reduction procedures. In Section \ref{sec:analysis}, we provide the detailed analysis of the data, including light curve, broad-band spectrum, index-luminosity relationship, and flares in X-rays. Then in Section \ref{sec:discuss} we discuss the implications of our results, where a possible explanation of flare differences among different LLAGNs is provided. A brief summary is devoted to the last section.
  
  \section{X-ray Observations} \label{sec:obs}
  
  \chandra\ provides high spatial and spectral resolution in soft X-ray band, which are necessary to diagnose the hot plasma around SMBH,
  and \nustar\ supplements crucial information in the hard X-ray band, which helps to investigate the innermost region of the BH in this LLAGN.
  Therefore we collected all the public archived \chandra\ and \nustar\ observations, 
  of which M81* is in the field of view, at the time of writing  this paper. No convincing short-term X-ray flare is detected during these
  observations, consistent with previous works \citep{2001MNRAS.321..767I, 2004ApJ...601..831L}.
  
  \subsection{\chandra} \label{subsec:chandra}
  
  \begin{table*}
    \caption{Observation Logs of M81* with HETG by \chandra}
    \label{tab:obs}
    \begin{tabular}{cccrccccl}
      \hline \hline
      OBSID &  Mission &  Instrument & Angle$^a$ & 
      Start Date &  Exposure$^b$ & $F_\mathrm{X}^c$ &
      Photon Index & Notes$^d$\\
      &   &  & \multicolumn{1}{|c|}{(\arcmin)}  &  & (ks) &
      $(10^{-11} \mathrm{erg~cm}^{-2} \mathrm{s}^{-1})$ & 
      & \\
      \hline
      6174 & \chandra & ACIS-S/HETG & \nodata & 2005-02-24 & 44.61& $1.08_{-0.03}^{+0.04}$   & $1.80_{-0.03}^{+0.03}$  & P \\
      6346 & \chandra & ACIS-S/HETG & \nodata & 2005-07-14 & 54.48& $1.02_{-0.03}^{+0.03}$   & $1.88_{-0.03}^{+0.03}$  & P \\
      6347 & \chandra & ACIS-S/HETG & \nodata & 2005-07-14 & 63.87& $1.06_{-0.03}^{+0.03}$   & $1.87_{-0.03}^{+0.03}$   & P \\
      5601 & \chandra & ACIS-S/HETG & \nodata & 2005-07-19 & 83.07& $1.13_{-0.03}^{+0.03}$   & $1.85_{-0.02}^{+0.02}$  & P \\
      5600 & \chandra & ACIS-S/HETG & \nodata & 2005-08-14 & 35.96& $0.90_{-0.03}^{+0.03}$   & $1.88_{-0.05}^{+0.03}$   & P \\
      6892 & \chandra & ACIS-S/HETG & \nodata & 2006-02-08 & 14.76& $0.87_{-0.05}^{+0.05}$   & $1.80_{-0.04}^{+0.08}$  & P \\
      6893 & \chandra & ACIS-S/HETG & \nodata & 2006-03-05 & 14.76& $0.96_{-0.06}^{+0.06}$   & $1.85_{-0.05}^{+0.07}$   & P \\
      6894 & \chandra & ACIS-S/HETG & \nodata & 2006-04-01 & 14.76& $0.96_{-0.06}^{+0.06}$   & $1.86_{-0.07}^{+0.06}$   & P \\
      6895 & \chandra & ACIS-S/HETG & \nodata & 2006-04-24 & 14.56& $0.94_{-0.05}^{+0.06}$   & $1.90_{-0.05}^{+0.08}$   & P \\
      6896 & \chandra & ACIS-S/HETG & \nodata & 2006-05-14 & 14.76& $0.95_{-0.06}^{+0.06}$   & $1.85_{-0.07}^{+0.05}$   & P \\
      6897 & \chandra & ACIS-S/HETG & \nodata & 2006-06-09 & 14.76& $0.90_{-0.05}^{+0.06}$   & $1.87_{-0.06}^{+0.06}$   & P \\
      6898 & \chandra & ACIS-S/HETG & \nodata & 2006-06-28 & 14.75& $0.94_{-0.06}^{+0.06}$   & $1.88_{-0.05}^{+0.07}$   & P \\
      6899 & \chandra & ACIS-S/HETG & \nodata & 2006-07-13 & 14.94& $1.04_{-0.06}^{+0.06}$   & $1.82_{-0.05}^{+0.07}$   &P \\
      6900 & \chandra & ACIS-S/HETG & \nodata & 2006-07-28 & 14.41& $1.87_{-0.08}^{+0.07}$   & $1.77_{-0.05}^{+0.04}$  & P \\
      6901 & \chandra & ACIS-S/HETG & \nodata & 2006-08-12 & 14.76& $1.96_{-0.08}^{+0.08}$   & $1.79_{-0.03}^{+0.07}$  & P \\
      20624 & \chandra & ACIS-S/HETG & \nodata & 2017-12-21 & 14.76& $1.77_{-0.08}^{+0.08}$  & $1.81_{-0.04}^{+0.06}$  & P \\
      \hline \hline
    \end{tabular}
    
    {\it Note}.
    The data are sorted by the start time of each observation. Errors are 90\% confidence level for one interesting parameter.
    \\
    $^a$  Nominal off-axis angle of the optical center of M81* on the detector. Blocks with no data mean on-axis ($<1\arcmin$). The distortion and broadness of  PSF within this separation are negligible.\\
    $^b$ Screened exposure time\\
    $^c$ Unabsorbed flux in 2 - 10 keV\\
    $^d$ Notes on zeroth order spectra. P means piled up, N means unpiled up, D means data are used for diffuse emission.  In this work we focus on the dispersed grating spectra, they do not suffer pile-up effects.
  \end{table*}

  \begin{table*}
    \caption{Observation Logs of M81* without gratings by \chandra}
    \label{tab:obs2}
    \begin{tabular}{cccrccccl}
      \hline \hline
      OBSID &  Mission &  Instrument & Angle$^a$ & 
      Start Date &  Exposure$^b$ & $F_\mathrm{X}^c$ &
      Photon Index & Notes$^d$\\
      &   &  & \multicolumn{1}{|c|}{(\arcmin)} &  & (ks) &
      $(10^{-11} \mathrm{erg~cm}^{-2} \mathrm{s}^{-1})$ & 
      & \\
      \hline
      390 & \chandra & ACIS-S & 1.78 & 2000-03-21 & 2.38   & \nodata & \nodata & P, \textemdash\ \\
      735 & \chandra & ACIS-S & 3.08 & 2000-05-07 & 50.02& $2.17_{-0.19}^{+0.20}$  & $1.83_{-0.06}^{+0.06}$   & P, S \\
      5935 & \chandra & ACIS-S & \nodata & 2005-05-26 & 10.78& $\Downarrow$ & $\Downarrow$ & P, S\\
      5936 & \chandra & ACIS-S & \nodata & 2005-05-28 & 11.41& $\Downarrow$ & $\Downarrow$ & P, S\\
      5937 & \chandra & ACIS-S & \nodata & 2005-06-01 & 12.01& $1.12_{-0.18}^{+0.20}$    & $1.75_{-0.12}^{+0.12}$   & P, S\\
      5938 & \chandra & ACIS-S & \nodata & 2005-06-03 & 11.81& $\Uparrow$ & $\Uparrow$ & P, S\\
      5939 & \chandra & ACIS-S & \nodata & 2005-06-06 & 11.61& $\Uparrow$ & $\Uparrow$ & P, S\\
      5940 & \chandra & ACIS-S & \nodata & 2005-06-09 & 11.98& $\Downarrow$ & $\Downarrow$ & P, S\\
      5941 & \chandra & ACIS-S & \nodata & 2005-06-11 & 11.81& $\Downarrow$ & $\Downarrow$ & P, S\\
      5942 & \chandra & ACIS-S & \nodata & 2005-06-15 & 11.95& $0.83_{-0.13}^{+0.14}$   & $1.87_{-0.12}^{+0.12}$   & P, S\\
      5943 & \chandra & ACIS-S & \nodata & 2005-06-18 & 12.01& $\Uparrow$ & $\Uparrow$ & P, S\\
      5944 & \chandra & ACIS-S & \nodata & 2005-06-21 & 11.81& $\Uparrow$ & $\Uparrow$ & P, S\\
      5945 & \chandra & ACIS-S & \nodata & 2005-06-24 & 10.40& $\Downarrow$ & $\Downarrow$ & P, S\\
      5946 & \chandra & ACIS-S & \nodata & 2005-06-26 & 11.82& $\Downarrow$ & $\Downarrow$ & P, S\\
      5947 & \chandra & ACIS-S & \nodata & 2005-06-29 & 10.70& $0.82_{-0.16}^{+0.18}$   & $1.84_{-0.14}^{+0.15}$   & P, S\\
      5948 & \chandra & ACIS-S & \nodata & 2005-07-03 & 12.03& $\Uparrow$ & $\Uparrow$ & P, S\\
      5949 & \chandra & ACIS-S & \nodata & 2005-07-06 & 12.02& $\Uparrow$ & $\Uparrow$ & P, S\\
      9805 & \chandra & ACIS-S &         1.21 & 2007-12-21 & 5.11& \nodata & \nodata & P, \textemdash\ \\
      9122 & \chandra & ACIS-S &         2.95 & 2008-02-01 & 9.91& \nodata & \nodata& P, \textemdash\\\
      9540 & \chandra & ACIS-S &      10.53 & 2008-08-24 & 25.52 & $0.84_{-0.02}^{+0.02}$   & $1.95_{-0.02}^{+0.02}$   & N\\  
      12301 & \chandra & ACIS-S &      2.59 & 2010-08-18 & 78.05& $1.37_{-0.18}^{+0.19}$  & $1.81_{-0.10}^{+0.10}$  & P, S\\
      18054 & \chandra & ACIS-I &       7.40 & 2016-06-21 & 65.71 & $1.47_{-0.02}^{+0.03}$   & $1.89_{-0.02}^{+0.02}$  & M\\  
      18875 & \chandra & ACIS-I &       7.40 & 2016-06-24 & 32.36 & $1.67_{-0.04}^{+0.04}$  & $1.85_{-0.02}^{+0.03}$  & M\\ 
      18047 & \chandra & ACIS-I &       7.19 & 2016-07-04 & 99.64 & $1.23_{-0.01}^{+0.01}$  & $1.76_{-0.01}^{+0.01}$   & M\\ 
      18048 & \chandra & ACIS-I &       6.63 & 2017-01-08 & 69.15 & $1.12_{-0.01}^{+0.01}$   & $1.74_{-0.02}^{+0.01}$  & M\\  
      19981 & \chandra & ACIS-I &       6.63 & 2017-01-11 & 29.64 & $1.15_{-0.02}^{+0.02}$   & $1.71_{-0.02}^{+0.02}$   & M\\  
      19982 & \chandra & ACIS-I &       7.94 & 2017-01-12 & 66.90 & $1.21_{-0.02}^{+0.02}$   &  $1.77_{-0.01}^{+0.02}$  & M\\  
      18053 & \chandra & ACIS-I &       7.99 & 2017-01-15 & 29.59 & $1.22_{-0.02}^{+0.02}$    & $1.74_{-0.02}^{+0.02}$  & M\\  
      18051 & \chandra & ACIS-I &     11.00 & 2017-01-24 & 36.56 & $1.08_{-0.02}^{+0.02}$   & $1.93_{-0.02}^{+0.02}$   & N\\  
      19993 & \chandra & ACIS-I &     10.95 & 2017-01-26 & 24.16 & $0.92_{-0.02}^{+0.02}$   & $1.92_{-0.03}^{+0.03}$  & N\\    
      19992 & \chandra & ACIS-I &     10.98 & 2017-01-28 & 34.61 & $0.94_{-0.02}^{+0.02}$   &  $1.90_{-0.02}^{+0.03}$  & N\\    
      \hline \hline
    \end{tabular}
    
    {\it Note}.
    The data are sorted by the start time of each observation. Errors are 90\% confidence level for one interesting parameter.\\
    $^a$ Nominal off-axis angle of the optical center of M81* on the detector, blocks with no data mean on-axis.
    The distortion and broadness of PSF within this separation are negligible.\\
    $^b$ Screened exposure time.\\
    $^c$ Unabsorbed flux in 2 - 10 keV, the arrows in this column and next column indicate the parameter values from the combined spectra.\\
    $^d$ P means high piled up, M means moderate piled up as discussed in \S \ref{subsec:chandra},  N means unpiled up, S means streak readout spectrum used,  \textemdash\ means observations with short exposure time which are not included in the analysis.
  \end{table*}
  
  M81* has been monitored regularly by \chandra. Among all the 47 observations publicly available, 16 times are observed with 
  the High Energy Transmission Grating (HETG) Spectrometer (Table \ref{tab:obs}) and the rest 31 times without it (Table \ref{tab:obs2}).
  The absorbed soft X-ray flux of M81* is above $10^{-11}~\mathrm{erg~cm}^{-2}~\mathrm{s}^{-1}$ over time, this will cause, in the Advanced CCD Imaging Spectrometer array (ACIS) of \chandra, a significant pile-up effect \citep{2001ApJ...562..575D}. Pile-up occurs when two or more photons land on the same or adjacent detector pixel within the same detector readout frame, and subsequently are either read as a single event with the summed energy or are discarded as a non-X-ray event.
  
  During the observations with HETG Spectrometer, due to dispersing spectra of gratings, the pile-up effect is negligible. Thus we only consider $\pm$1st order
  spectra of the medium energy gratings (MEGs) and the high-energy gratings (HEGs) for M81*. The total exposure time of filtered data is 0.44 Ms.
  Although the nucleus region of zeroth order spectra suffers pile-up effect, we took advantage of the high spatial resolution of \chandra\ 
  on-axis imaging  to analyze the iron lines in surrounding diffuse emission.
  
  On the other hand, for the observations without grating, on-axis ones are also used for analyzing the diffuse iron lines in virtue of the larger effective area. There are some observations that M81* is not the main target thus has a large off-axis location. We found that due to the broadness of point spread function (PSF) and the vignetting, pile-up effect is much suppressed in this case. Although the nearby point sources are blended into the PSF of M81*, their impact is found to be small. We added up the luminosity of these nearby sources (\citealt{2003ApJS..144..213S,2011ApJ...735...26S} , see Figure 1 in \citealt{2007ApJ...669..830Y} for the distribution of near-nuclear point sources), which is about $5\times 10^{38}~\mathrm{erg}~\mathrm{s}^{-1}$ in 0.5 - 8 keV. This is 50 times lower than that of  M81*, thus we do not consider their contamination below.
  
  We reprocessed all the \chandra\ data with {\sc ciao} v4.10 - 4.11 and {\tt\string deflare} routine was implemented to remove the 
  background flare intervals. Spectroscopic analyses were performed using {\sc isis} v1.62. 
  We accessed the pileup fractions of large off-axis angle observations by {\tt\string pileup\_map} tool in {\sc ciao}. 
  Although most of them show a fraction 7\% - 18\%, we argue that this tool is the indicator for on-axis sources, this fraction should be
  less for large off-axis sources. More sophisticated simulations with {\sc SAOTrace} and {\sc MARX} can help to refine it, but it's beyond the scope of this work.
  Thus, we use symbol M in Table \ref{tab:obs2} to  represent the moderate piled up. The unpiled up observations are estimated to be less than 5\% fraction.
  
  For HETG data, X-ray flux and photon index are estimated from $\pm$1st order spectra. 
  For the imaging data, although high piled-up spectra of on-axis observations can be analyzed under the pileup model \citep{2001ApJ...562..575D}, the model parameters can not be constrained well for individual observations due to low signal-to-noise (S/N) ratio. Thus we do not list their parameters in Table \ref{tab:obs2}. Instead, for on-axis observations with long exposure, spectra from the ACIS readout streak are analyzed instead.  Because the exposure time of the available readout streak is about 100 times shorter than the on-axis spectrum, the S/N ratio is low in 15 observations (OBSID 5935 - 5949, individually). Thus we combined every 5 streak spectra together according to these intermittent observations which were proceeded in 6 weeks. Furthermore, pileup model also wasn't implemented for off-axis observations due to inaccuracy in this mode\footnote{The \chandra\ ABC Guide to Pileup, \url{http://cxc.harvard.edu/ciao/download/doc/pileup\_abc.pdf}}.
  And they are estimated by extracting from an adequate region to enclose the whole PSF for large off-axis observations.
  For imaging observations, backgrounds are extracted from nearby source-free regions.
  
  In {\sc isis}, we combined $\pm$1st order spectra and restricted wavelength range to 1.7 - 28\AA\ for HETG data.
  For imaging observations, energy range 0.5 - 8.0 keV is noticed analogously. A minimum S/N ratio 5 or 3 in each 
  spectra bin was adopted.
  
  A simple model, {\tt\string TBabs $\times $ powerlaw}, is applied separately to each observation, and {\tt\string cflux} is used to get the desired flux.
  If residuals are significant in Fe emission region (6-7 keV), a {\tt\string gaussian} model will be added. Actually, the emission lines are fairly weak compared to the continuum, the most significant lines are Fe K$\alpha$, Fe {\sc xxv} and Fe {\sc xxvi} whose total flux is about 1\% of the 2 - 10 keV continuum \citep{2004A&A...422...77P, 2007ApJ...669..830Y,2018MNRAS.476.5698Y}.
  The absorption model {\tt\string TBabs} \citep{2000ApJ...542..914W} with the cross sections of \citet{1996ApJ...465..487V}. For absorbed {\tt\string powerlaw} model, there exists a weak anti-correlation between
  the column density and the photon index.
  Thus in order to avoid this anti-correlation and compare with the results from other missions,
  the line of sight (LOS) column density $\nh$ was fixed to that of Milky Way, i.e. 
  $\nh = 4.2\times10^{20}\mathrm{cm}^{-2}$ \citep{1990ARA&A..28..215D}.
  Note that the low variation of $\nh$ in M81* (see Sec.\ \ref{sec:intro}) leads to insignificant impact on the flux above 2 keV. The unabsorbed 2-10 keV flux of M81* is measured with convolution model {\tt\string cflux}. 
  
  \subsection{\nustar} \label{subsec:nustar}
  M81* has been observed three times by \nustar\, using the Focal Plane Modules (FPMA and FPMB) (Table \ref{tab:obs3}).
  \nustar\ has a sufficient high temporal resolution, thus does not suffer pile-up effect. We follow standard pipelines, 
  {\tt\string nupipeline} and {\tt\string nuproducts} in {\sc heasoft} version 6.27 - 6.28, to extract products. 
  Besides removing the South Atlantic Anomaly (SAA) intervals from exposure time, we additionally take filtering for the
  ``tentacle'' region\footnote{\nustar\ FAQ 36, \url{https://heasarc.gsfc.nasa.gov/docs/nustar/nustar_faq.html\#saa3}}. Without this filtering, there exist two $\sim$ 1 hr flare features in hard X-ray band during OBSID 60101049002 (hereafter Nu15). The data of OBSID 50401006001 is excluded from our analysis due to short exposure. During OBSID 50401006002 (hereafter Nu18), M81* was off-axis and had a distorted PSF shape. 
  However, in hard X-rays there are no other detectable sources within 3\arcmin\ circular region around the nucleus, so we use an adequate size of  region files to enclose the whole PSF of M81* to extract the spectra. Although we cannot exclude possible contribution from the stellar bulge, the X-ray emission is dominated by nucleus of M81 and portion increases with increasing photon energy \citep{2003ApJS..144..213S}. Background counts are taken from source-free circular regions in the field of view. 
  
  The spectra of FPMA and FPMB are then combined together in {\sc isis}. Following the methods reported in previous section, we can derive the model parameters, as reported in Table \ref{tab:obs3}. All the cleaned products in this work have been extracted after the barycenter correction by each corresponding tool.

  \begin{table*}
    \caption{Observation Logs of M81* by \nustar}
    \label{tab:obs3}
    \begin{tabular}{cccccccl}
      \hline \hline
      OBSID &  Mission &  Instrument & Start Date &  Exposure$^a$ & $F_\mathrm{X}^b$ & Photon Index & Notes\\
      &   &  &  & (ks) & $(10^{-11} \mathrm{erg~cm}^{-2} \mathrm{s}^{-1})$ & & \\
      \hline
      60101049002 & \nustar & FPMA/FPMB & 2015-05-18 &   202.8 & $2.24_{-0.01}^{+0.01}$   & $1.89_{-0.01}^{+0.01}$  & on-axis\\ 
      50401006002 & \nustar & FPMA/FPMB & 2018-09-06 &      88.0 & $2.16_{-0.02}^{+0.02}$   & $1.95_{-0.03}^{+0.03}$ & off-axis, $\sim5\arcmin$\\
      50401006001 & \nustar & FPMA/FPMB & 2018-09-06 & 0.5 & \nodata & \nodata &  short exposure\\
      \hline \hline
    \end{tabular}
    
    {\it Notes}.
    The data are sorted by the start time of each observation. Errors are 90\% confidence level for one interesting parameter.\\
    $^a$ Screened exposure time, averaged by each module.\\
    $^b$ Unabsorbed flux in 2 - 10 keV.
  \end{table*}
  
  \begin{figure}
    \includegraphics[width=\columnwidth]{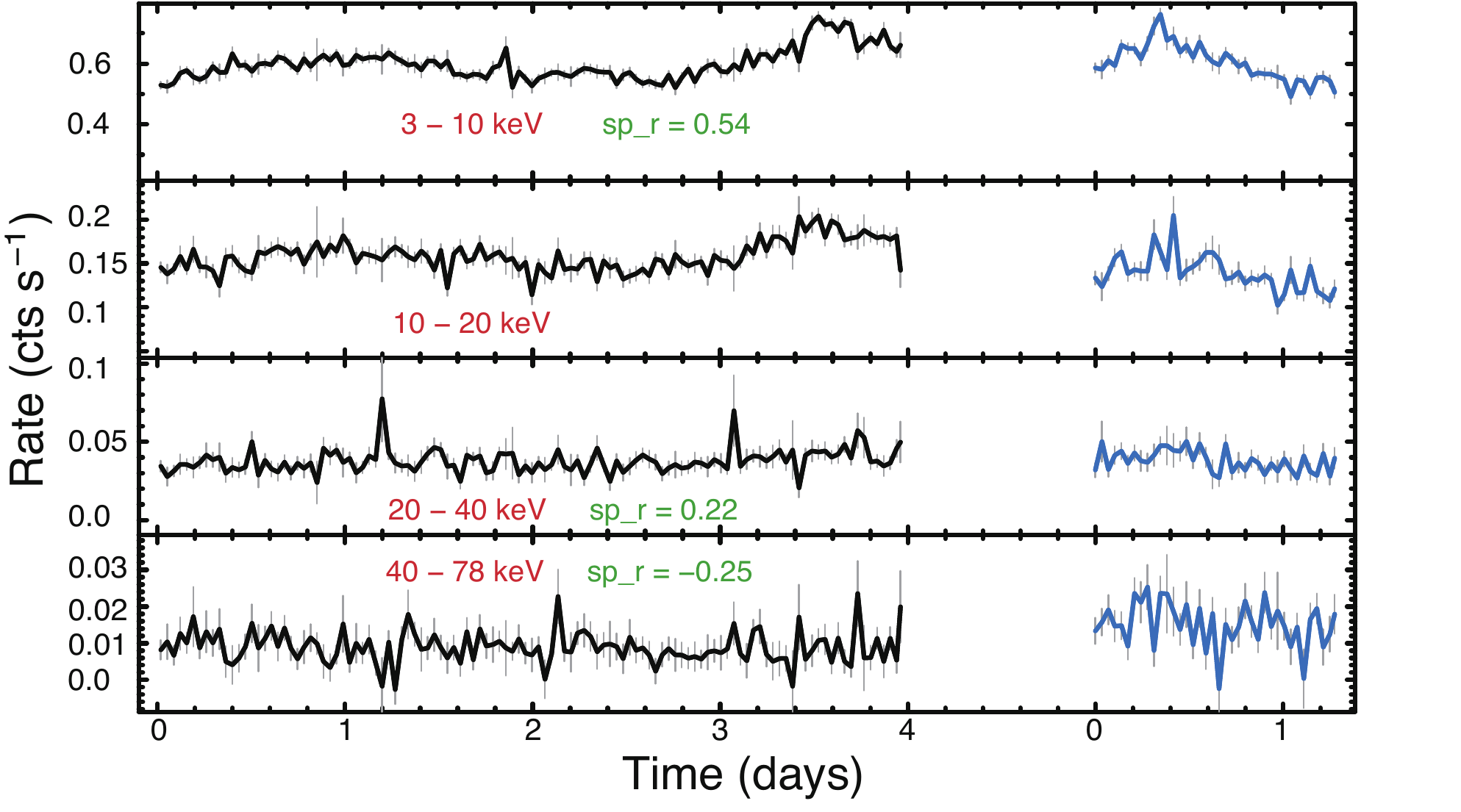}
    \caption{The light curves of M81* in bins of 3000 s with \nustar. The count rates are the mean values between the FPMA and FPMB modules, corrected for livetime, PSF losses and vignetting. The background is also subtracted. Black and blue lines refer to OBSID 60101049002 and 50401006002. Each panel shows the Spearman's rank  correlation coefficient, $sp_r$, between the count rates of each energy range and of 10 – 20 keV.
      Note that the time separation between two observations is shortened for clarity.  The
      fluctuations of the high energy band are likely to be due to the low count rate, moreover, 
      these short-term intervals are less than the light crossing time for the M81* case.
    }\label{fig:lc}
  \end{figure}
  
  \begin{figure}
    \includegraphics[width=0.95\columnwidth]{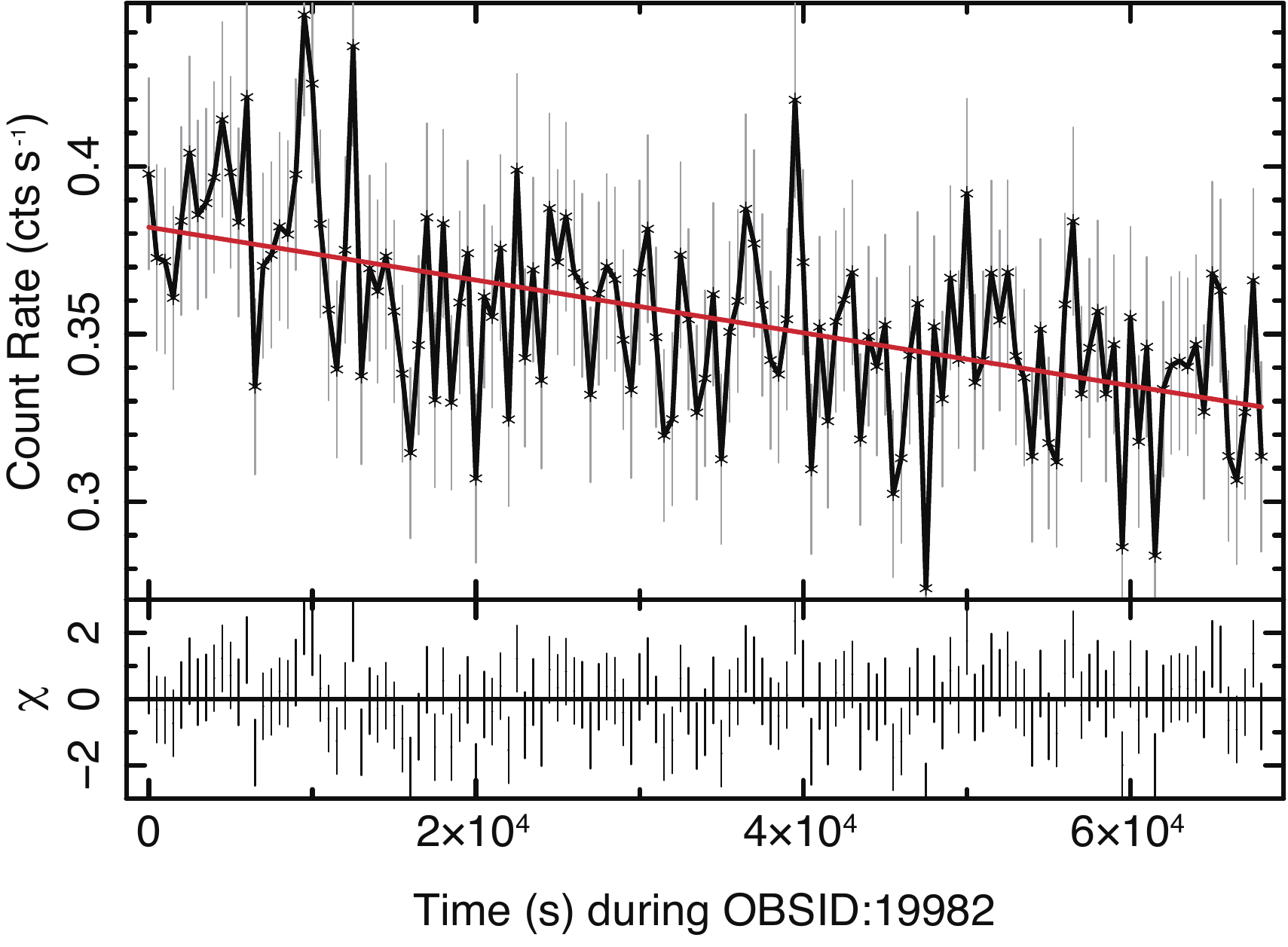}
    \caption{The light curves of M81* in bins of  500 s in 2 - 8 keV band with \chandra\ during OBSID 19982. The red solid line is the best linear fit to the light curve.}\label{fig:19982}
  \end{figure}
  
  \section{Data Analysis} \label{sec:analysis}
  
  \subsection{light curves in X-rays}
  With duration of 100-200 ks (cf. Table \ref{tab:obs3}), the \nustar\ observations provide the light curves in timescale of $\sim 10$ hrs. This is shown in Figure \ref{fig:lc}, where we consider energy bands in  3 - 10, 10 - 20, 20 - 40 and 40 - 78 keV, respectively, for panels from top to bottom. The data are binned to 3000s.
  
  Several results can be derived immediately from Figure \ref{fig:lc}. First, as illustrated by the upper two panels, emission below 20 keV in M81* shows significant, $\sim 20-30 \%$, variations on $\sim 40$ ks timescale (see also \citealt{2018MNRAS.476.5698Y}). Considering the light crossing timescale around a $M_{\rm BH}=7\times10^{7}\,M_{\sun}$ BH, this implies that the variation should come from a compact $\lesssim110 R_{\rm s}$ region.\footnote{We emphasize that, contribution from point sources in circum-nuclear region is weak compared to M81*, and no bright transients have been reported \citep{2011ApJ...735...26S}. Besides, although the PSFs of \nustar\ and \chandra\ (in off-axis mode) are large, the spatial distribution shape coincides with a central point source. This also suggest that the variation observed should come from M81* itself.} On the other hand, the hard X-ray emission above 20 keV is fairly steady. On longer timescales, a moderate variability is observed in 2-10 keV. The 2-10 keV flux of M81* can vary by a factor of $\sim 3$ at different epochs spanning $\sim$17 yrs, as shown in Table \ref{tab:obs2}. However, there seems no clear global trend (e.g., decline or increase) among different epochs.
  
  Second, no flares can be detected (at S/N$>$3) among the two \nustar\ observations. The apparent flare features in the hard X-ray light-curves (above 20 keV, the lower two panels of Figure \ref{fig:lc}) are likely due to fluctuations (thus unreal), since their significance is below $2\sigma$ detection in M81*.
  The lack-of-flare property in X-rays is confirmed and further strengthened by all the 47 \chandra\ observations between 2000 and 2017 \citep{2007ApJ...669..830Y,2010ApJ...720.1033M}, where the exposure time of each observation varies from $\sim 2$ ks to $\sim 100$ ks (see Tables \ref{tab:obs2} and \ref{tab:obs3}).
  
  However, we found that a few off-axis \chandra\ observations (e.g., OBSID 19982) have similar long-term gradual variations as shown in \nustar\ observation, its flux almost linearly decreases over 15\% in 67 ks as shown in Figure \ref{fig:19982}. 
  
  Third, emissions below and above 20 keV seem to have weak connections. For this investigation, we cross-correlated light curves at different energy bands to that of 10-20 keV, where we take Spearman's rank coefficient $sp_r$ to measure the significance of their correlations. As labelled in each panel of Figure \ref{fig:lc}, only the 3-10 keV band has a clear but weak positive correlation ($sp_r=0.54$) to the 10-20 keV band; the hard X-rays above 20 keV, on the other hand, are consistent with null correlations ($sp_r\sim$0.2).
  
  \begin{figure}
    \includegraphics[width=\columnwidth]{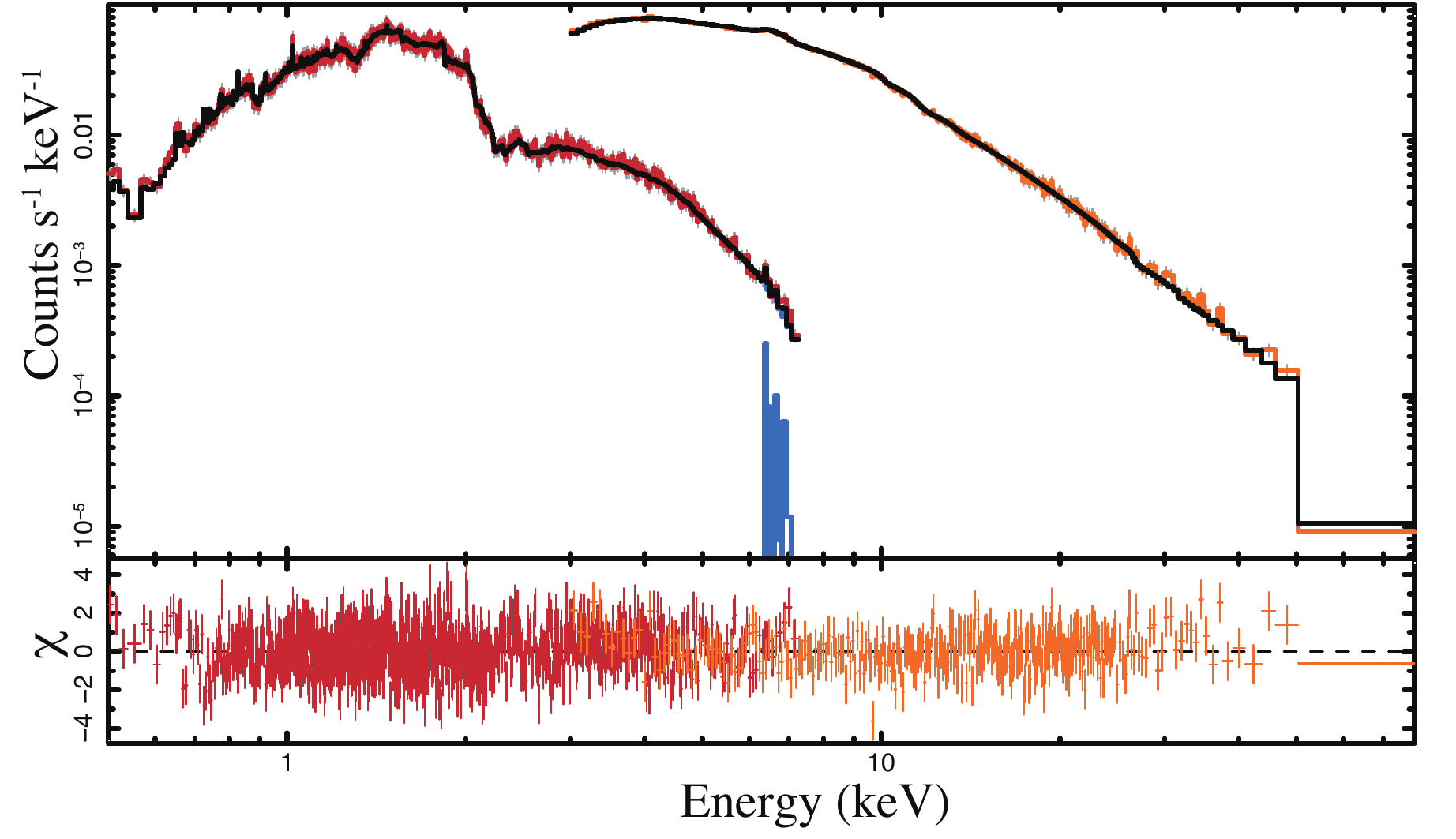}
    \caption{M81* \chandra\ $\pm$1st HETG (red line) and FPMA/FPMB combined \nustar(orange line) spectra, together with the best-fit cutoff powerlaw model, plus 3 gaussian lines representing the Fe lines around 6.7 - 7.0 keV.}
    \label{fig:spec}
  \end{figure}
  
  \subsection{broadband X-ray spectrum}
  We now explore the broadband spectral properties of M81*. For this investigation, we utilize both \chandra/HETG data and the two long-exposure \nustar\ (0.29 Ms in total) data.
  
  Although the origin of the X-ray continuum remains inclusive (e.g, \citealt{2008ApJ...681..905M, 2010ApJ...720.1033M, 2014MNRAS.438.2804N, 2017ApJ...836..104X}), the spectral shape of M81* in X-rays is somewhat easy to model. 
  The Compton hump is found to be absent or weak, from a joint fit of the \suzaku\ and \nustar\ observations of M81* \citep{2003A&A...400..145P, 2018MNRAS.476.5698Y}. In this spirit, we joint fit the broadband X-ray emission of \chandra\ and \nustar\ under a simple phenomenological {\tt\string TBabs $\times$ cutoffpl} model, where a free cross-normalization constant of \nustar\ spectra relative to \chandra/HETG is included. Here the spectrum of {\tt\string cutoffpl} takes the form $F_\epsilon\propto \epsilon^{1-\Gamma}\exp{(-\epsilon/E_{\rm c})}$. Modeling based on alternative models can be found in \citet{2018MNRAS.476.5698Y}. Again, in our work the absorption is fixed to $N_{H} = 4.2\times10^{20}\mathrm{cm}^{-2}$. The results are shown in Figure \ref{fig:spec}. We find that $\Gamma = 1.85\pm0.01$ and $E_{\rm c} = 192.82^{+32.67}_{-12.24}$ keV. 
  We note that our value of $E_{\rm c}$ is lower than that ($E_{\rm c} \sim 250$ keV) reported in \citet{2018MNRAS.476.5698Y}, due to differences in model setup.
  
  \begin{figure}
    \includegraphics[width=\columnwidth]{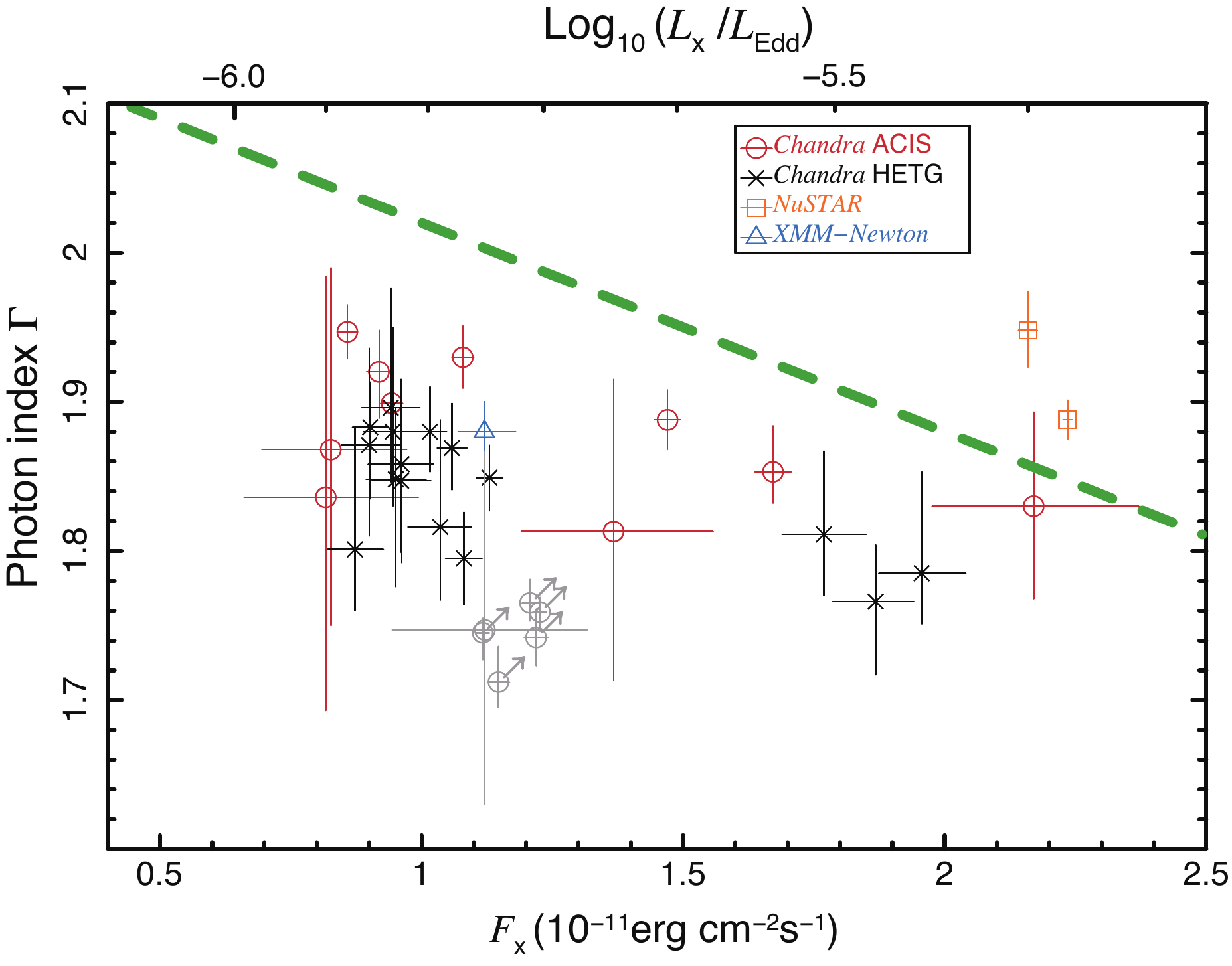}
    \caption{Relationships between the photon index and X-ray flux in 2 - 10 keV.
      The abscissa is the logarithm of the X-ray Eddington ratio.
      The green dashed line represents the best fit linear anti-correlation from \textit{Swift} observations.
      The blue triangle for \xmm\ is derived from the literature \citep{2004A&A...422...77P}.
      The data points from off-axis spectra and readout streak spectra with \chandra\ imaging observations (see Section \ref{subsec:chandra})  are marked both as circles. Among them, 5 gray data points with arrows suggest the possible intrinsic data of the moderated pileup observations (see Section \ref{sec:diagram}).
    }\label{fig:idxflx}
  \end{figure}
  
  \subsection{The index-luminosity relation in X-rays}\label{sec:diagram}
  We now explore the correlation between the X-ray power-law photon index and the X-ray luminosity. Here we take the 2 - 10 keV X-ray luminosity as representative. The index-luminosity correlation can be used to probe the accretion physics, where a ``V''-shaped relationship is observed (\citealt{2015MNRAS.447.1692Y}, and references therein), i.e. a negative (also called as ``harder when brighter'') relation for LLAGNs (e.g., \citealt{2016MNRAS.459.3963C} for individual sources), and a positive (``softer when brighter'') relation for bright AGNs. The index-luminosity relationship in M81* is explored in literature, based on data of various X-ray missions \citep{1996PASJ...48..237I,2004ApJ...601..831L,2010ApJ...720.1033M,2016MNRAS.459.3963C}.
  
  Figure \ref{fig:idxflx} shows a plot of photon index against the X-ray luminosity (and the Eddington ratio, defined as $L_{\rm X}(2-10)/L_{\rm Edd}$). The red circles and black crosses show, respectively, data from \chandra/ACIS and \chandra/HETG. We caution that, the 5 data points from \chandra\ off-axis observations (OBSID 8047, 18048, 19981, 19982 and 18053, shown by gray circles),  with flux $1.1\sim1.3\times10^{-11} \mathrm{erg~cm}^{-2} \mathrm{s}^{-1}$ and $\Gamma\approx 1.7$-1.8, have highest pileup fraction in our dataset. The pileup effect will trigger more signals of hard photon, \footnote{see introduction section of \url{https://xmmweb.esac.esa.int/docs/documents/CAL-TN-0214-214-1.pdf}}, so the intrinsic photon index should be softer that that measured. Since it is difficult to quantify this effect, we do not include these data for further analysis. The orange squares are data from \nustar\ and the blue triangle is from \xmm\  \citep{2004A&A...422...77P}.
  
  From this plot, at given luminosity, we observe a scatter of $\Delta\Gamma\approx 0.1-0.2$ in $\Gamma$. We further observe, if the two \nustar\ data points are excluded, a negative correlation between $\Gamma$ and X-ray luminosity (and flux). The Spearman's rank correlation coefficient is -0.46. Note that, as shown by a green dashed line, a similar anti-correlation is also observed in \textit{Swift} observations, where a flux-binned spectral modeling is adopted \citep{2016MNRAS.459.3963C}. Clearly, there is a systematic offset between \textit{Swift} data and our data, i.e. \textit{Swift} observations tend to have a larger $\Gamma$ value (equivalently, softer in X-rays) at given $L_{\rm X}$. The $\Gamma$ of \nustar\ also tends to be larger  at given luminosity.\footnote{The photon index of M81* is almost unchanged even when the hard X-ray band is included because its X-ray continuum can be well described by a  power-law \citep{2018MNRAS.476.5698Y,2021NatAs...5..928S}.} This is probably because of contamination in soft X-rays from stellar bulge regions \citep{2003ApJS..144..213S}.
  
  \begin{figure}
    \includegraphics[width=\columnwidth]{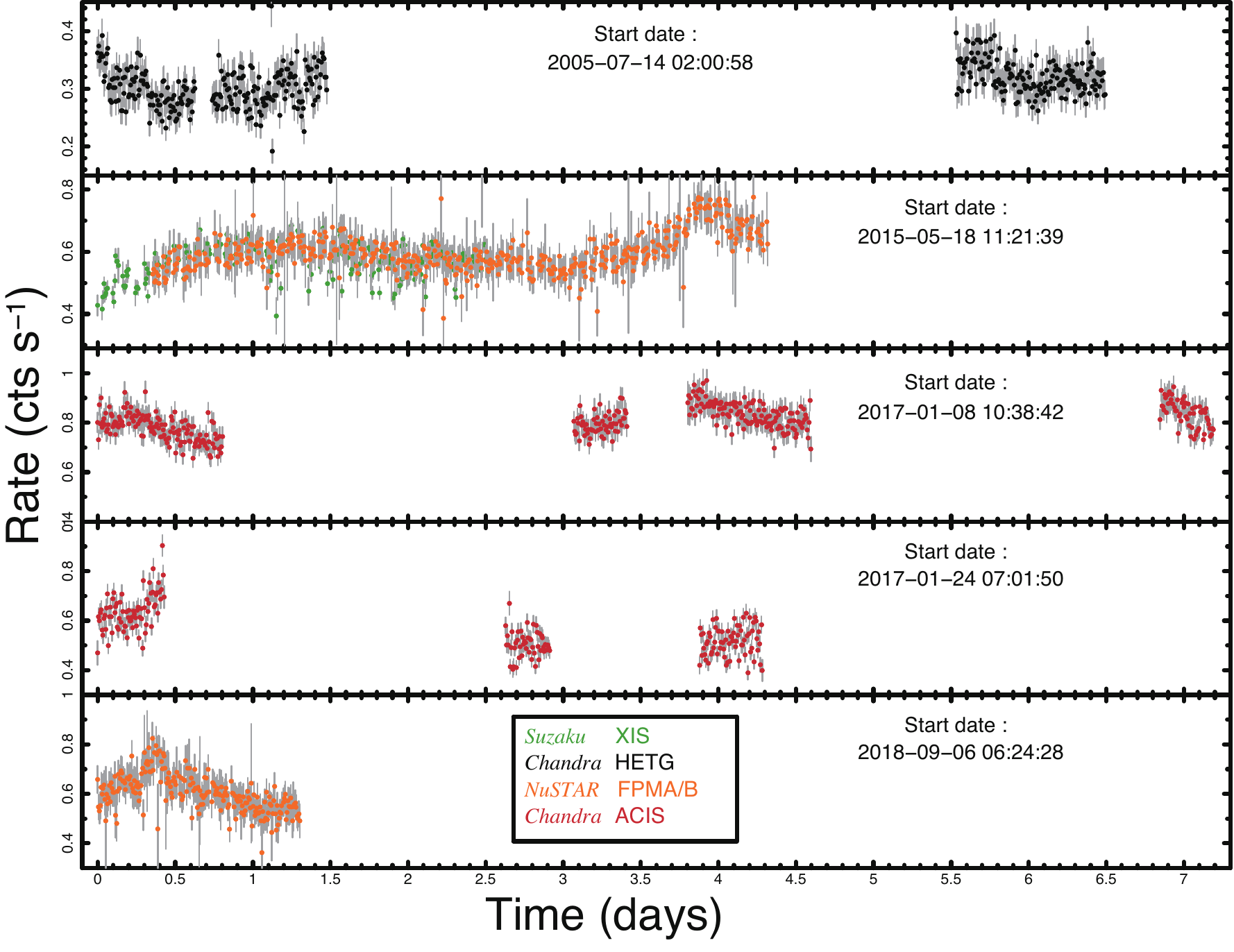}
    \caption{Light curves of M81* in bins of  500s. The black points represent combined data from \chandra\ HETG $\pm$1st, $\pm$2nd, $\pm$3rd orders in 1.7 - 28 \AA, the orange points are from \nustar\ in 3 - 10 keV band, and the red points are from \chandra\ ACIS-I  in 0.5 - 8 keV band. The green points are scaled data from \suzaku\ XISs in 0.5 - 8 keV. The start date of each light curve is labelled in each panel. The corresponding OBSID is listed in Tables  \ref{tab:obs}, \ref{tab:obs2}, and \ref{tab:obs3}. Note that normalization may be different between panels 3 and 4 (shown by red points), where they have different off-axis angles and suffer moderate pile-up effect.}
    \label{fig:all_lc}
  \end{figure}
  
  \subsection{weak and broad flares in X-rays}\label{sec:flare}
  
  M81* is observed to be fairly steady on decade-timescale (i.e., from 2005 to 2014, \citealt{2016MNRAS.459.3963C}). Meanwhile, as shown in Figure \ref{fig:lc}, it also does not show large ($>$2) amplitude variations on timescales of hours to days. Here, we test if there are any weak flares in M81*, with the motivation to compare with observations in Sgr A* (See Section \ref{sec:flare_model}). Since the dynamical timescale, i.e. the shortest timescale of an accretion system, at $3 R_{\rm s}$ of M81* is $t_{\rm dyn} \equiv R/V_{\rm k}(R)\approx 5\times10^{3}$ s (here $V_{\rm k}=(GM_{\rm BH}/R)^{1/2}$ is the Keplerian velocity and $t_{\rm dyn}\propto R^{3/2}$), we only explore variations on timescales greater than several hours. Besides, as suggested by Figure \ref{fig:lc} (see \citealt{2001MNRAS.321..767I, 2004ApJ...601..831L, 2016NatPh..12..772K}, and Sec.\ \ref{sec:flare_model} below), we limit ourselves to X-rays at low energy bands, where we take $\lesssim$10 keV.
  
  \begin{figure}
    \includegraphics[width=\columnwidth]{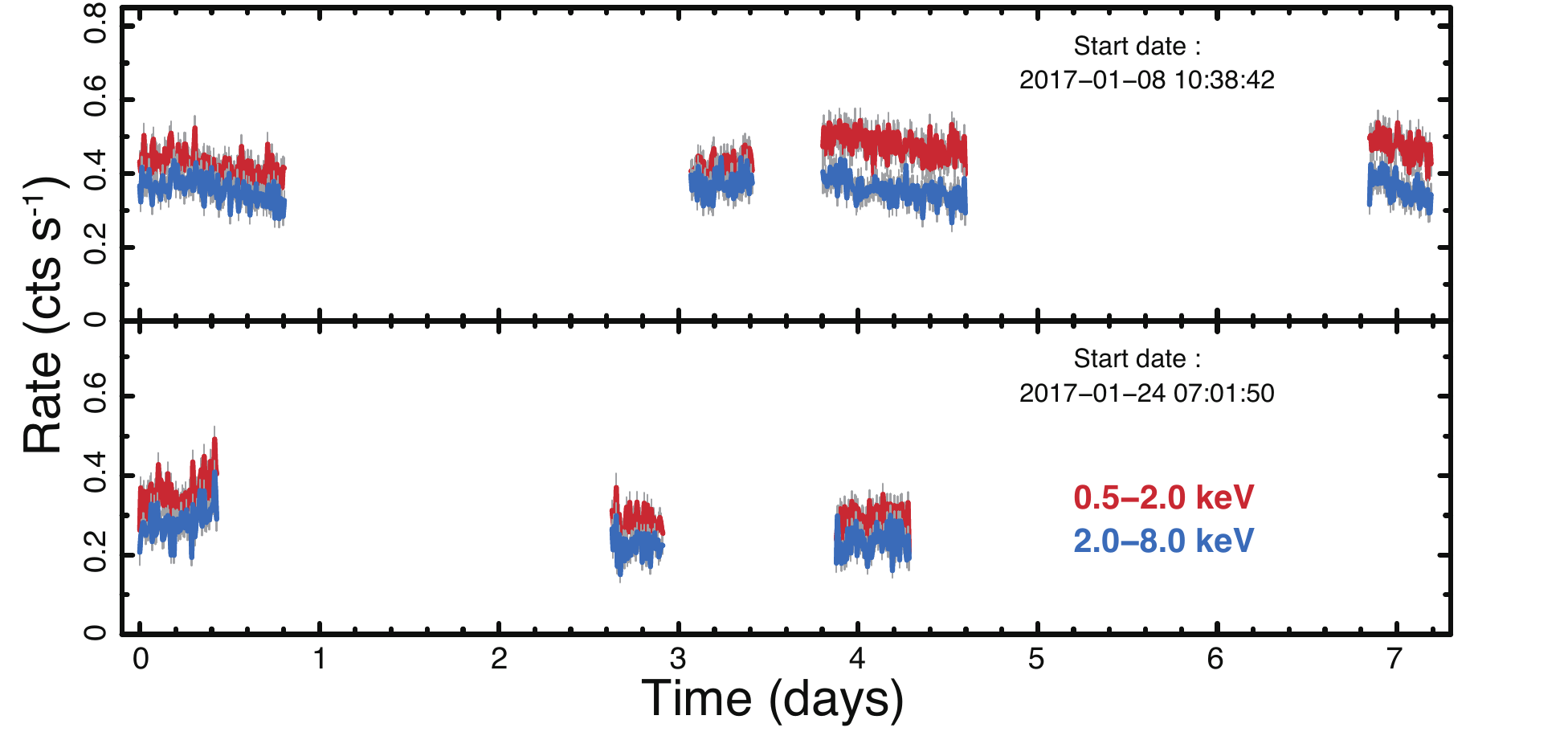}
    \caption{Light curves of two bands from 7 off-axis ACIS-I M81* observations in Figure \ref{fig:all_lc} are shown in bins of 500s. Note that the effective areas will decrease with different rates as off-axis angles increase in different energy bands, so there is a modest discrepancy in effective area ratios between the top and bottom panels.}
    \label{fig:svsh}
  \end{figure}
  
  \subsubsection{Test 1: weak flares in soft X-rays}
  We checked the archived X-ray observations (including \xmm, \rxte) in past two decades. The exposure time of most observations is short, thus it is challenging to test variability on hour-day timescales. Here we first focus on the 2015 \nustar\ observation that has a 202.8 ks exposure time (Nu15 hereafter, see Table \ref{tab:obs3}). As shown in Figure \ref{fig:lc}, there is evidence of a weak and broad flare in 3-10 keV band. To highlight the significance, we rebinned the data to 500s, and the results are shown by orange points in the second panel (from top to bottom) of Figure \ref{fig:all_lc}. We also include a simultaneous \suzaku\ observation (OBSID 710017010; \citealt{2018MNRAS.476.5698Y}) that precedes the \nustar\ data. The green points in the second panel shows the scaled count rate of averaged 3 XISs in 0.5$-$8 keV of \suzaku\, to that of \nustar\ with overlap time. The two light curves match each other well, and we hereafter refer this combined data as NS15.
  
  We take a simple {\tt\string constant+gaussian} model to fit the light curve in 3-10 keV. The {\tt\string gaussian} component takes the form $F_{\rm peak}\exp(-(t-t_{\rm peak})^2/\sigma_{\rm flare}^2)$. We find that, besides the stable $\approx 0.577\pm 0.004\,{\rm cts\, s}^{-1}$ emission in 90\% confidence level, there exists a flare that has a peak flux of $F_{\rm peak}\approx 0.152\pm 0.017\,{\rm cts\, s}^{-1}$ and timescale of $\sigma_{\rm flare}\approx 0.22\pm 0.03\,{\rm d}$. If this flare is similar to that of  \citet{2016NatPh..12..772K}, the amplitude of the flare may be more significant in soft X-rays.
  
  \subsubsection{Test 2: quasi-period oscillation in soft X-rays}
  Below we consider an alternative interpretation. Considering there may be two local maxima within the $\sim$4d observation of NS15, we may further aggressively interpret the soft X-rays of NS15 as a sinusoidal-like variation. Additional hints of such variation is also observed by data with shorter exposure times. From example, the \xmm\ observation of M81* also reveals a decreasing trend in  $\sim1.4$ days (\citealt{2004A&A...422...77P}, see also Figure\ \ref{fig:19982}). Figure \ref{fig:all_lc} shows all the long-duration X-ray observations. For completeness, we also include adjacent observations that have short gaps. These data are shown in the rest of the panels of Figure \ref{fig:all_lc}, where the start time is labelled in each panel. Interestingly, despite large gaps between epochs, from the third panel of Figure \ref{fig:all_lc} we find that the four epoch off-axis \chandra\ data also suggest a sinusoidal-like variation pattern.
  
  We explore if there exists a quasi-periodicity variation in the soft X-rays. Because of the low variation (only weak and broad flares exist), we find that $z$-transformed cross-correlation algorithm (ZDCF, \citealt{1997ASSL..218..163A}), which is an interpolation method 
  for discrete sparse astronomical time series, is not suitable for our investigation. Generally, we cannot predict a convincing phase lag between NS15 and the \chandra\ data in 2017 January (see also \citealt{2006ApJ...652.1531M}).
  
  Instead, we adopt the minimum string-length method \citep{1983MNRAS.203..917D}, which can establish the period from a relatively small number of randomly spaced observations over a long span of time. The period is chosen at the minimum of the string-length $SL$ (see \citealt{1983MNRAS.203..917D} for details), which is defined as
  \begin{equation}
  SL=\sum_{i=1}^{n}\left[\left(m_{i}-m_{i-1}\right)^{2}+\left(\phi_{i}-\phi_{i-1}\right)^{2}\right]^{1 / 2}. \label{eq:sl}
  \end{equation}
  Here  $m$ and $\phi$ are the normalized amplitude and phase of the X-ray emission if we assume there is a periodic variation (of assumed period $P$) in the signal. The normalization assures that, both the amplitude and the phase are re-scaled as dimensionless quantities, with their values restricted to the -0.25 - 0.25 range \citep{1983MNRAS.203..917D}. All the data are then folded into $n$ bins. Note that technically when $0$-th bin refers to the last $n$-th bin.
  From equation \ref{eq:sl},  SL measures the sum of both amplitude and phase. When the assumed period $P$ matches the real period of the data, a minimum value of $SL$ will then be derived. We note that, the absolute value of $SL$ has no meaning, since it depends on the observational sampling, the number of bins $n$, and the intrinsic periodicity $P$.
  
\begin{figure}
    \includegraphics[width=1.15\columnwidth]{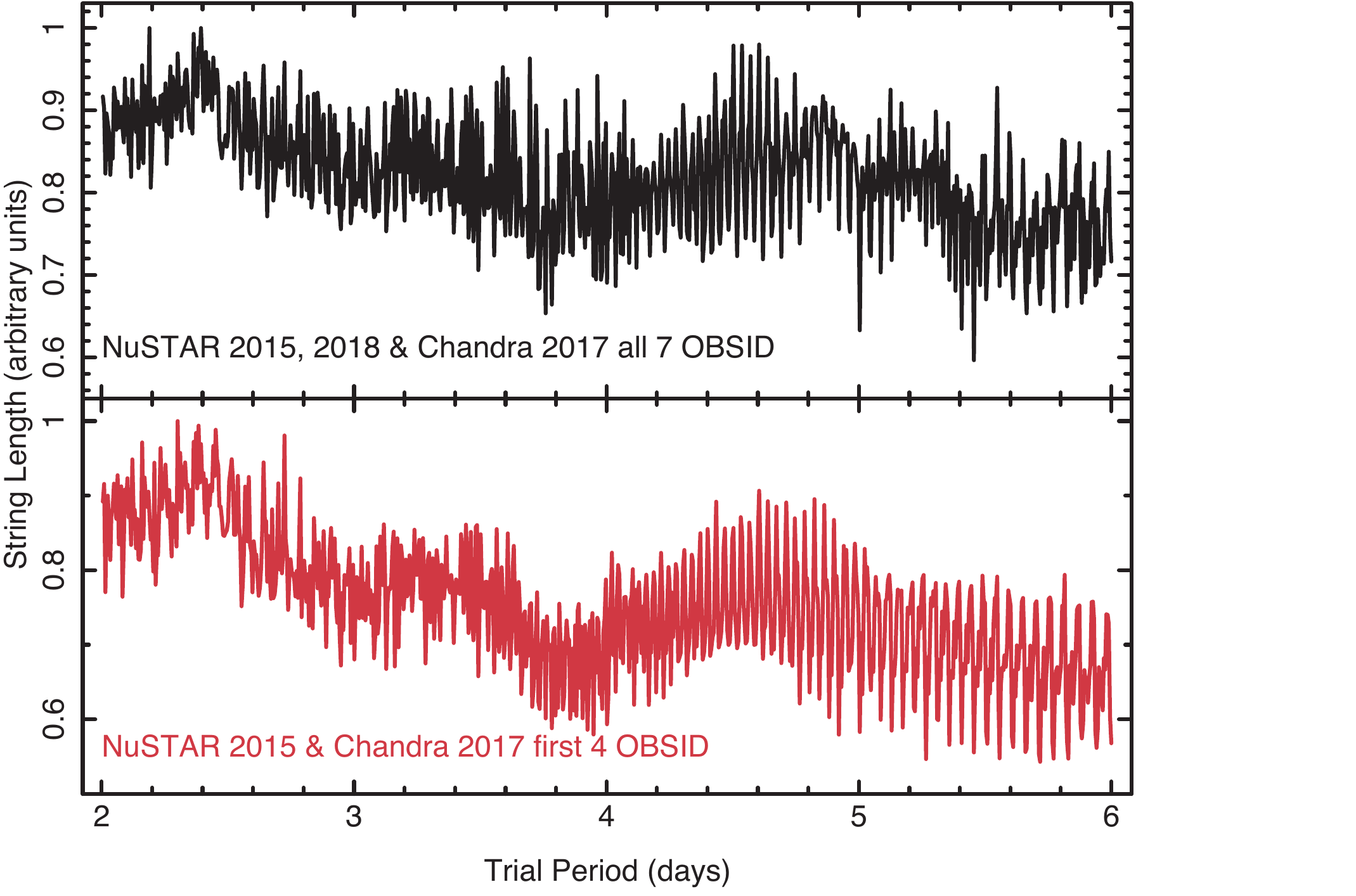}
    \caption{String length ($SL$, see Equation \ref{eq:sl}) against trial period. {\it Lower Panel}: only data of NS15 and the first 4 \chandra\ observations in 2017. {\it Upper Panel}: additionally include Nu18 and the rest 3 \chandra\ observations in 2017. A minimum at period $P\sim 3.8$ d is clearly detected in the lower panel, and it is much broadened in the upper panel.}
    \label{fig:slm}
  \end{figure}

  In our analysis, we first consider NS15 and the first four epochs of \chandra\ only (i.e., the second and third panels of Figure \ref{fig:all_lc}), and the string-length at different trial periods is shown in the lower panel of Figure \ref{fig:slm}. From this plot, a best-value period of $3.85_{-0.02}^{+0.02}$ days (at 90\% confidence level) is observed. Although another local minimum above 5 days is also presented, we consider it as a harmonic feature. Besides, due to limitations in the exposure of each observation, it is difficult to constrain such long quasi-period. Moreover, the shape near the minimum is fairly broad, suggesting that the flares are only quasi-periodic. 
  
  We then additionally include data of Nu18 and the rest three \chandra\ observations in 2017, and the result is shown in the upper panel of Figure \ref{fig:slm}. The significance of the minimum $SL$ at $\sim$3.8d is now much weakened.
  
  \begin{figure}
    \includegraphics[width=\columnwidth]{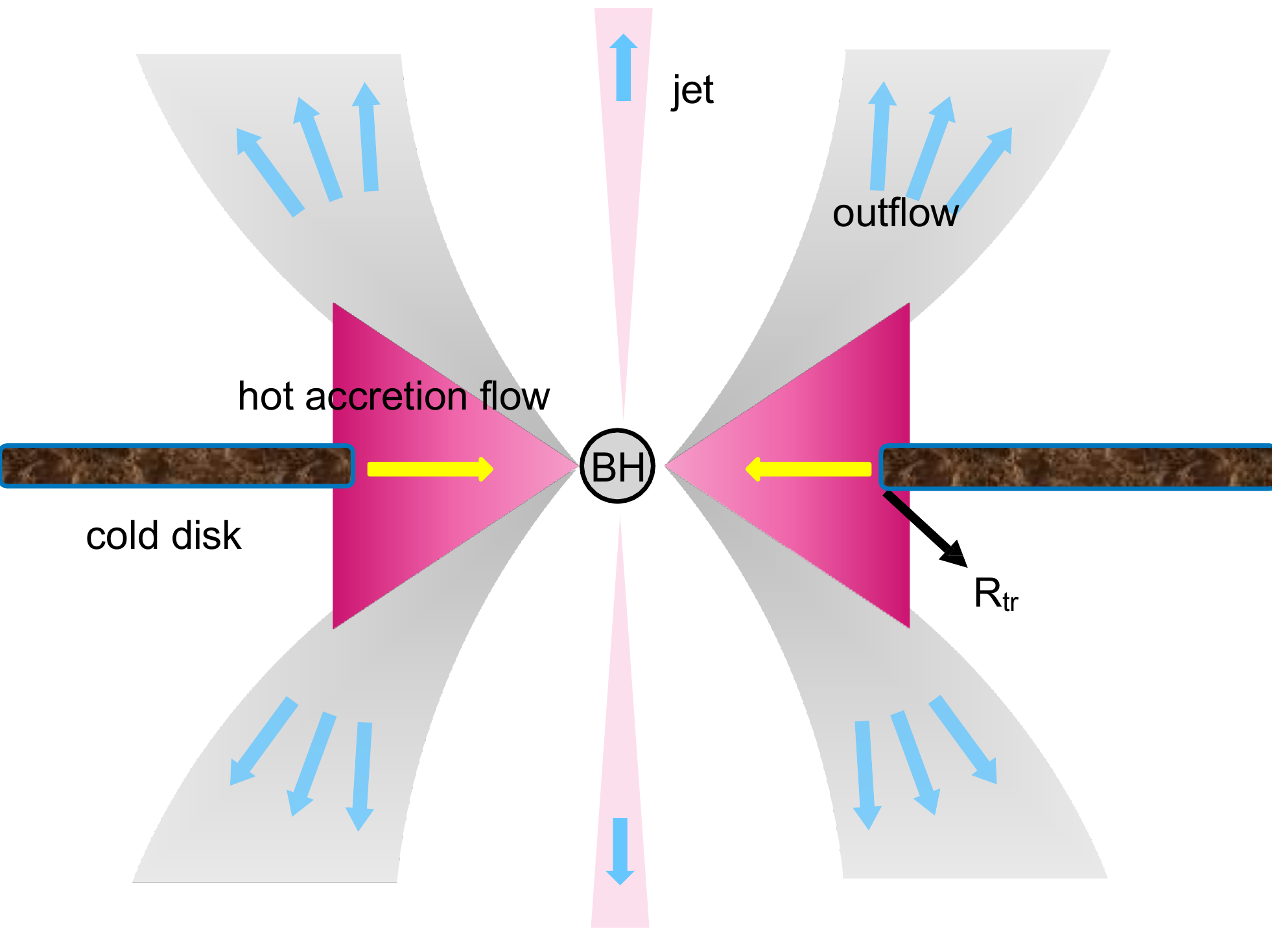}
    \caption{Schematic picture of the accretion in M81*. There exists a cold disk (show as dark gray region) that is truncated at $R_{\rm tr}\approx 300 R_{\rm s}$. Inside $R_{\rm tr}$ is a hot accretion flow (purple region), which also launches hot outflow from its surface. Relativistic jet (light pink region) perpendicular to accretion flow is launched from a region close to BH.}
    \label{fig:m81schem}
  \end{figure}
  
  \section{Discussions}\label{sec:discuss}
  
  We now discuss the theoretical implication of M81* observations. A comparison to properties of a much fainter system Sgr A* is also addressed.
  
  \subsection{general picture of M81*}
  Figure \ref{fig:m81schem} shows a schematic picture of our understanding of accretion in LLAGNs \citep[see also][]{2014ARA&A..52..529Y}. In this hot accretion-jet model, there exists a cold geometrically-thin disk (show as dark gray region, \citealt{1973A&A....24..337S}) that is truncated at $R_{\rm tr}\approx 300 R_{\rm s}$. Inside $R_{\rm tr}$ there exist a hot accretion flow (purple region, \citealt{1994ApJ...428L..13N}), which also launches hot outflow from its surface. A relativistic jet (light pink region) perpendicular to accretion flow is launched from a region close to BH. We find that this picture provides a comprehensive explanation of the observations of M81*, whose $L_{\rm bol}/L_{\rm Edd}\sim 10^{-5}$.
  
  It is known that in hot accretion flow the truncation radius $R_{\rm tr}$ decreases with increasing accretion rate \citep[and bolometric luminosity,][]{2004ApJ...612..724Y}. With $L_{\rm bol}/L_{\rm Edd}\sim 10^{-5}$, it is reasonable to have $R_{\rm tr}\sim 10^{2-3}\,R_{\rm s}$ in M81*. Indeed, M81* shows clear evidence that there exists a cold disk outside of $\sim 300 R_{\rm s}$ \citep[e.g., ][]{1999ApJ...525L..89Q, 2007ApJ...671..118D, 2009RAA.....9..401X}.
  
  There are several consequences of this value of $R_{\rm tr}$. First, the large $R_{\rm tr}$ here indicates that, the opening solid angle of the cold disk to that of hard X-rays (generated at inner regions of hot accretion flow, i.e. $\lesssim 10-20 R_{\rm s}$) will also be small. This agrees with the observational constraint of a non-detection of the Compton reflection hump (reflection parameter $R<0.11$) in X-rays \citep{2003A&A...400..145P, 2018MNRAS.476.5698Y}. 
  
  Second, because accretion gas outside of $R_{\rm tr}$ is in the form of a cold disk, we should not expect to have strong and spatially-extended (up to $10^{5-6} R_s$) bremsstrahlung emission in X-rays, a prominent component that is observed in two much fainter systems \citep{2003ApJ...591..891B, 2006MNRAS.372...21A, 2013Sci...341..981W, 2017MNRAS.466.1477R}, i.e. Sgr A* ($L_{\rm bol}/L_{\rm Edd}\approx 1-2\times10^{-9}$, \citealt{2003ApJ...591..891B, 2019ApJ...881L...2B, 2019MNRAS.483.5614M, 2023ApJ...942...20X}) and M87 ($L_{\rm bol}/L_{\rm Edd}\approx 1\times10^{-6}$, \citealt{2016MNRAS.457.3801P, 2021ApJ...911L..11E}). In these two systems, bremsstrahlung emission are produced by hot gas from a large region, from $10^{3-4}\, R_s$ to a radius beyond the Bondi radius ($\sim 10^{5}\, R_s$).
  
  Third, the origin of continuum emission in X-rays. Our understanding to this is advanced thanks to \nustar\ observations. From a non-physical phenomenological fit of \suzaku\ and \nustar\ data under a cutoff power-law model, no apparent curvature can be observed within the \nustar\ band \citep{2018MNRAS.476.5698Y}. In this case, modelling the X-ray emission of M81* by the bremstrahlung model requires the gas temperature being greater than 100 keV (modelling result not shown here). This is unphysical, since electrons at such high temperature ($k T_e/m_e c^2\sim 0.2$) should produce strong radiation through inverse Compton scattering process. In the hot accretion-jet model of LLAGN, the most promising mechanism to produce X-rays inside $\lesssim R_{\rm tr}\sim 10^{2-3} R_s$ of a $L_{\rm bol}/L_{\rm Edd}\sim 10^{-5}$ system is a combination of inverse Compton scattering (dominant) and bremstrahlung (secondary, and actually negligible) by hot electrons. At the accretion rate of M81*, a powerlaw spectrum of the Comptonization component is totally expected. This is consistent with observations (e.g., \chandra:  \citealt{2003ApJS..144..213S}, \nustar: \citealt{2018MNRAS.476.5698Y}).
  
  There also exist a weak but extended X-ray emission outside of $R_{\rm tr}$ in M81* \citep{2003ApJS..144..213S}. For this component, we note that hot accretion flow can generate strong outflow that can propagate distant away from the BH, as shown in Figure \ref{fig:m81schem}. Numerical simulations of hot accretion flow suggest that these outflows are much hotter and more dilute (because of higher velocity) than gases within the hot accretion flow itself if exists \citep{2021NatAs...5..928S}. Consequently, bremsstrahlung from these hot but dilute outflowing gas may be responsible for the weak diffuse/extended X-ray emission observed in \chandra\ \citep{2003A&A...400..145P}.\footnote{The emissivity of bremsstrahlung, $j_\nu({\rm bremss}) \propto n_e^2\,T^{-1/2}$ \citep{2013LNP...873.....G}, decreases with decreasing gas density and increasing gas temperature.} 
  
  Fourth, the emission lines in X-rays. Weak Fe K$\alpha$, Fe {\sc xxv} and Fe {\sc xxvi} emission (not absorption) lines are detected in X-rays (e.g., \citealt{2003A&A...400..145P, 2021NatAs...5..928S}). The Fe K$\alpha$ line of neutral or weakly ionized iron may originate from reflection of a truncated accretion disk (or a torus of cold gas) that are irradiated by the central LLAGN \citep{2018MNRAS.476.5698Y}. Indeed, if the observed Fe K$\alpha$ line originates in reflection of Compton-thick material, a much higher reflection fraction, $R\sim 0.3-0.4$, should be measured based on the $\sim$40 eV EW of the Fe K$\alpha$ line \citep[e.g.][]{1991MNRAS.249..352G}. 
  
  On the other hand, Fe {\sc xxv} and Fe {\sc xxvi} may originate from outflows in M81* \citep{2021NatAs...5..928S}, where they are collisionally ionized. The existence of outflow in LLAGN is indeed a theoretical expectation \citep[][and references therein]{2012ApJ...761..130Y,2015ApJ...804..101Y}. Observationally a receding velocity $\sim 2000-3000~{\rm km~s^{-1}}$ is detected clearly only in Fe {\sc xxvi} \citep{2004A&A...422...77P, 2007ApJ...669..830Y}. In the outflow-origin interpretation, assuming that M81* has a low inclination, both approaching and receding components can be observed in the X-ray spectrum (if observations have sufficient S/N), but the approaching one may be weakened due to asymmetry below and above the equatorial plane of the hot accretion flow \citep{2021NatAs...5..928S}.

  \begin{table*}
    \caption{Observational X-ray Flare Properties Among Sgr A*, M31* and M81*}
    \label{tab:flares}
    \begin{tabular}{ccccccccc}
      \hline \hline
      Name & $M_{\rm BH}$ & $L_{\rm bol}$  & $\lambda_{\rm Edd}$ &  \multicolumn{5}{c}{Observational X-ray Flare Properties}
      \\
      \cline{5-9}
      & ($M_{\sun}$) & ($\rm erg\,s^{-1}$) &  ($L_{\rm bol}/L_{\rm Edd}$) & occurrence rate 
      & duration & amplitude$^a$  & spectral change$^b$ & Refs.$^c$\\
      \hline
      Sgr A*& $4 \times10^6$    & $10^{36}$                   & $\sim10^{-9}$ &  several per day     & $\sim$ 1 hour  & $10^{\sim 0-2}$ & harder & B03, N13\\
      M81*  & $7 \times10^7$   & $2.1\times10^{41}$ & $\sim10^{-5}$ & rare & $\sim$ 1 day  & weak, $\lesssim$2 & stable & K16, this work \\ 
      M31* &$1.4 \times 10^8$ & $\sim 10^{38}$         &$\sim 10^{-8}$ & $\sim$1-2 per year$^d$ & hours to days & $10^{\sim 0-1}$ & harder & G10, L11\\
      \hline \hline
    \end{tabular}

    $^a$ the amplitude is measured as the ratio of the flux at the peak of flare ($L_{\rm flare}^{\rm peak}$) to the flux of non-flare state ($L_{\rm non.flare}$).\\    
    $^b$ spectral differences to that of the non-flare period.\\
    $^c$ References: B03 - \citet{2003ApJ...591..891B}, N13 - \citet{2013ApJ...774...42N}, K16 - \citet{2016NatPh..12..772K}, G10 - \citet{2010ApJ...710..755G}, L11 - \citet{2011ApJ...728L..10L}.\\
    $^d$ The occurrence rate here only applies to the outburst period \citep{2011ApJ...728L..10L}.
  \end{table*}
  
  \subsection{flares (or outbursts) in hot accretion flow: comparison to Sgr A* and M31*} \label{sec:flare_model}
  
  In this section, we discuss the flares (or outbursts when the amplitude is high) in hot accretion flows. We summarize the flare properties of several LLAGNs, i.e. Sgr A*, M31* and M81*, in Table \ref{tab:flares}. Besides the BH mass and bolometric Eddington ratio ($\lambda_{\rm Edd}=L_{\rm bol}/L_{\rm Edd}$), we include in the table the occurrence rate, typical duration and amplitude (w.r.t. non-flare state) of each flare, and spectral change (flare state, w.r.t. non-flare state).
  
  \subsubsection{basic properties of flares in LLAGNs}
  Observationally, strong flares are known to be rare in LLAGNs like M81  \citep[e.g.][and Sec. \ref{sec:flare}]{2008ApJ...681..905M, 2016NatPh..12..772K}, M31* ($L_{\rm X}\approx 10^{-10} L_{\rm Edd}$ with $M_{\rm BH}\sim 10^8\,M_{\sun}$, e.g., \citealt{2010ApJ...710..755G, 2011ApJ...728L..10L}), and M87. 
  
  We notice that, based on high-resolution radio observation \citet{2016NatPh..12..772K} detected a discrete knot/blob moving at a mildly relativistic velocity away from BH of M81*. Associated with this knot ejection, they also find in soft ($<$2 keV) X-rays a flare that precedes the radio flare by about 12 days. On the other hand, there is no clear variation in hard X-rays above 2 keV. Besides, \citet{2016NatPh..12..772K} also carried out systematical analysis of all the archival \swift\ data of M81 and found this event to be unique in six years (2006-2011).
  
  On the other hand, Sgr A*, with $L_{\rm X}\approx 10^{-11} L_{\rm Edd}$ and $M_{\rm BH}\approx 4\times10^6\,M_{\sun}$, is unique in that, while it spends most of its time in quiescent state, it exhibits numerous flares in radio, submillimeter, infrared and X-rays (e.g., \citealt{2001Natur.413...45B,2003Natur.425..934G,2006ApJ...650..189Y,2009ApJ...702..178Y,2011ApJ...728...37D,2015MNRAS.453..172P,2018ApJ...863...15W}). The flare properties 
  have been studied in detail in Sgr A*, e.g., based on simultaneous multi-band monitoring \citet{2009ApJ...702..178Y} and statistical analysis \citep{2013ApJ...774...42N, 2015ApJ...810...19L}. Their occurrence rate is a couple times per day and the typical duration of each flare is about an hour 
  (e.g., \citealt{2013ApJ...774...42N, 2016MNRAS.456.1438Y, 2018MNRAS.473..306Y,2018ApJ...863...15W}). The flares are most evident in infrared and X-rays (e.g., 2-10 keV), where the flux can increase by a factor of $\ga 10$ within $\sim 10$ minutes (e.g., \citealt{2003Natur.425..934G,2013ApJ...774...42N}). Besides, the flares are observed to be linear polarized in NIR and submm and have a distinct spectral index in infrared and in X-rays (\citealt{2017ApJ...843...96Z, 2018ApJ...863...15W, 2019ApJ...886...96H}). The broadband emission in flares are believed to be of synchrotron origin (e.g., \citealt{2017ApJ...850L..30Y,2017MNRAS.468.2447P}).
  
  \subsubsection{model interpretation}
  Theoretically the flares in Sgr A* are interpreted as an abrupt energy releases of magnetic reconnection, where electrons can be accelerated to form the energy distribution as power-law, i.e. $n(\gamma)\propto \gamma^{-p}$. Consequently, it will produce synchrotron emission from radio upto X-rays (and even $\gamma$ rays). One promising model is the magnetohydynamic model, mimic to the corona-mass-ejection phenomenon in Sun (CME model, see e.g., \citealt{2009MNRAS.395.2183Y, 2015ApJ...810...19L, 2022ApJ...933...55C}; see also \citealt{2014MNRAS.441.1005D,2020MNRAS.494.4168D} for alternative but similar models). In this model, magnetic reconnection occurred at the surface of the hot accretion flow results in the formation of flux ropes, which are then ejected out. Energetic electrons accelerated in the current sheet flow into the flux rope region and their synchrotron radiation is responsible for the flares observed. The ejection (outward motion) and expansion of plasma blob can also explain the time lags among light curves in different wavebands.
  
  Below we try to apply this CME model to other LLAGN systems, with an emphasis on the difference in black hole mass and accretion rate among different systems. Because of marginal flare detections in M81*, we here limit ourselves to qualitative instead of quantitative comparisons.
  
  First the flare duration. In Sgr A*, statistically the flare lasts 0.5$-$8 ks \citep{2013ApJ...774...42N}. Assuming that the duration of each flare follows a linear dependence on $M_{\rm BH}$ (notice that $t_{\rm dyn}\propto M_{\rm BH} (R/R_{\rm s})^{3/2}$), it will be $\sim10-$160 ks in M81* and $\sim 20-$300 ks in M31*. Besides, among the 3Ms of observations of Sgr A*, only 39 X-ray flares are detected. Considering the duration of each flares, we may estimate an occupation fraction in time to be $\sim 1\%$. If we assume such time-occupation rate also applies to M81* and M31*, then detecting such long (hours to days) flare but with low time-occupation fraction requires high cadence monitoring, thus is admittedly challenging.
  
  Second the flare amplitude. For each flare, the total  magnetic power ($E_{\rm b}$) restored before the eruption can be crudely estimated to be proportional to accretion rate $\dot{M}$, and $\dot{M}\propto \dot{m}M_{\rm BH}$. Here $\dot{m}$ defines the accretion rate in Eddington unit $\dot{M}_{\rm Edd}$.\footnote{The Eddington accretion rate is defined as $\dot{M}_{\rm Edd} = L_{\rm Edd}/(\eta c^2)$. The radiative efficiency $\eta$ of a hot accretion flow depends on the details of accretion physics \citep{2012MNRAS.427.1580X} and black hole spin as well \citep{2012MNRAS.420.1195N}. Here for simplicity a typical value of $\eta=10\%$ is adopted.} Then, the luminosity of the flare $L_{\rm flare}$, crudely estimated as $\propto E_{\rm b}/t_{\rm dyn}$, will be at least insensitive to BH mass, i.e., $L_{\rm flare}\propto M_{\rm BH}^{\sim 0}$. Consequently, we should observe a decrease in $L_{\rm flare}/L_{\rm non.flare}$ with increasing BH mass, which is consistent with the comparison between Sgr A* and M81*. Indeed, bright flares ($\gtrsim 10$x the quiescent flux) are never observed in M81*, and the only clearly-detected flare as reported in \citet{2016NatPh..12..772K} has $L_{\rm flare}^{\rm peak}/L_{\rm non.flare}\lesssim 2$ in soft X-rays, while it can reach $L_{\rm flare}^{\rm peak}/L_{\rm non.flare}\sim 100-$400 in Sgr A* \citep{2013ApJ...774...42N, 2017ApJ...843...96Z}.
  
  Third, the spectral property during flare state. For simplicity, we assume the power-law distribution of electrons accelerated by magnetic reconnection, $p$, is the same among different systems. Because the synchrotron cooling timescale of electrons with energy $\gamma m_{e}c^{2}$ is $t_{\rm syn}(\gamma)\propto \gamma^{-1} B^{-1}$ and the magnetic field strength scales as $B\propto \dot{M}^{1/2}\propto \dot{m}^{1/2}M_{\rm BH}^{1/2}$, we expect to have, in systems with high $\dot{m}$ and $M_{\rm BH}$, much shorter cooling timescale, thus the fraction of energetic (i.e., large-$\gamma$) electrons will be significantly reduced. In this case, the model predicts that the flare remains invisible/undetectable in systems with high $\dot{m}$ and $M_{\rm BH}$. This agrees with observations that, in Sgr A* the flares can be clearly detected in both soft and hard X-rays, while in M81* they can only be observed in soft X-rays \citep{2016NatPh..12..772K}.
  
  Finally, we note that, the CME model can also produce a QPO light curve, where the QPO period relates to the (rotational) orbital period of the moving blob \citep{2023MNRAS.520.1271L}.
  
  \section{Summary} \label{sec:summary}
  In this work, we have systematically analyzed the archived \chandra\ and \nustar\ X-ray data of M81*. We provided a detailed study of the light curves and statistics of spectral parameters. Our main results can be summarized as follows:
  \begin{enumerate}
    \item The light curves of \nustar\ observations show variation in $3-10$ keV X-ray band, but are quite stable in $>10$ keV band. No correlation can be found among different X-ray bands.
    \item The 2$-$10 keV X-ray flux and photon index show a similar anti-correlation trend as observed with \textit{Swift}, which is typically observed in LLAGNs and is characteristic in hot accretion flows like ADAF.
    \item The X-ray continuum of M81* can be interpreted as dominated by inverse Compton scattering in hot accretion flow.
    \item A variation on $\sim$ 2 - 5 days timescales in soft X-ray band is detected. Several simple models are tested, i.e. isolate flares, and sinusoidal-like ones (also supported by available long-exposure observations). Besides, due to limited exposure time, current data can not confirm the quasi-periodicity in soft X-rays of M81*.
    \item A possible explanation of this variation is a magneto-hydrodynamic model of ``coronal-mass-ejection'' phenomenon from the hot accretion flow in M81*. When differences in both the accretion rate and BH mass are considered, the flares in other LLAGNs, including Sgr A* and M31*, can also be understood naturally.
  \end{enumerate}
  
  \section*{Acknowledgements}
  F.G.X. is supported in part by National SKA Program of China (No. 2020SKA0110102), the National Natural Science Foundation of China (NSFC Nos. 12192220 and 12192223), and the Youth Innovation Promotion Association of CAS (Y202064).
  
  \section*{Data Availability}
  
  The correlated \chandra\ and \nustar\ raw data used in this paper are available at CXC\footnote{\url{https://cda.cfa.harvard.edu/chaser/}} 
  and HEASARC\footnote{\url{https://heasarc.gsfc.nasa.gov/docs/archive.html}} website.
  The reduced data underlying this paper will be shared on reasonable
  request to the corresponding author.
  
  
  
  \bibliographystyle{mnras}
  \bibliography{m81_main} 

\begin{thebibliography}{}
\makeatletter
\relax
\def\mn@urlcharsother{\let\do\@makeother \do\$\do\&\do\#\do\^\do\_\do\%\do\~}
\def\mn@doi{\begingroup\mn@urlcharsother \@ifnextchar [ {\mn@doi@}
  {\mn@doi@[]}}
\def\mn@doi@[#1]#2{\def\@tempa{#1}\ifx\@tempa\@empty \href
  {http://dx.doi.org/#2} {doi:#2}\else \href {http://dx.doi.org/#2} {#1}\fi
  \endgroup}
\def\mn@eprint#1#2{\mn@eprint@#1:#2::\@nil}
\def\mn@eprint@arXiv#1{\href {http://arxiv.org/abs/#1} {{\tt arXiv:#1}}}
\def\mn@eprint@dblp#1{\href {http://dblp.uni-trier.de/rec/bibtex/#1.xml}
  {dblp:#1}}
\def\mn@eprint@#1:#2:#3:#4\@nil{\def\@tempa {#1}\def\@tempb {#2}\def\@tempc
  {#3}\ifx \@tempc \@empty \let \@tempc \@tempb \let \@tempb \@tempa \fi \ifx
  \@tempb \@empty \def\@tempb {arXiv}\fi \@ifundefined
  {mn@eprint@\@tempb}{\@tempb:\@tempc}{\expandafter \expandafter \csname
  mn@eprint@\@tempb\endcsname \expandafter{\@tempc}}}

\bibitem[\protect\citeauthoryear{{Alexander}}{{Alexander}}{1997}]{1997ASSL..218..163A}
{Alexander} T.,  1997, {Is AGN Variability Correlated with Other AGN
  Properties? ZDCF Analysis of Small Samples of Sparse Light Curves}.
Springer Netherlands, p.~163, \mn@doi{10.1007/978-94-015-8941-3_14}

\bibitem[\protect\citeauthoryear{{Allen}, {Dunn}, {Fabian}, {Taylor}  \&
  {Reynolds}}{{Allen} et~al.}{2006}]{2006MNRAS.372...21A}
{Allen} S.~W.,  {Dunn} R.~J.~H.,  {Fabian} A.~C.,  {Taylor} G.~B.,   {Reynolds}
  C.~S.,  2006, \mn@doi [\mnras] {10.1111/j.1365-2966.2006.10778.x}, \href
  {https://ui.adsabs.harvard.edu/abs/2006MNRAS.372...21A} {372, 21}

\bibitem[\protect\citeauthoryear{{Baganoff} et~al.,}{{Baganoff}
  et~al.}{2001}]{2001Natur.413...45B}
{Baganoff} F.~K.,  et~al., 2001, \mn@doi [\nat] {10.1038/35092510}, \href
  {https://ui.adsabs.harvard.edu/abs/2001Natur.413...45B} {413, 45}

\bibitem[\protect\citeauthoryear{{Baganoff} et~al.,}{{Baganoff}
  et~al.}{2003}]{2003ApJ...591..891B}
{Baganoff} F.~K.,  et~al., 2003, \mn@doi [\apj] {10.1086/375145}, \href
  {https://ui.adsabs.harvard.edu/abs/2003ApJ...591..891B} {591, 891}

\bibitem[\protect\citeauthoryear{{Bietenholz}, {Bartel}  \&
  {Rupen}}{{Bietenholz} et~al.}{2000}]{2000ApJ...532..895B}
{Bietenholz} M.~F.,  {Bartel} N.,   {Rupen} M.~P.,  2000, \mn@doi [\apj]
  {10.1086/308623}, \href
  {https://ui.adsabs.harvard.edu/abs/2000ApJ...532..895B} {532, 895}

\bibitem[\protect\citeauthoryear{{Bower} et~al.,}{{Bower}
  et~al.}{2019}]{2019ApJ...881L...2B}
{Bower} G.~C.,  et~al., 2019, \mn@doi [\apjl] {10.3847/2041-8213/ab3397}, \href
  {https://ui.adsabs.harvard.edu/abs/2019ApJ...881L...2B} {881, L2}

\bibitem[\protect\citeauthoryear{{Connolly}, {McHardy}, {Skipper}  \&
  {Emmanoulopoulos}}{{Connolly} et~al.}{2016}]{2016MNRAS.459.3963C}
{Connolly} S.~D.,  {McHardy} I.~M.,  {Skipper} C.~J.,   {Emmanoulopoulos} D.,
  2016, \mn@doi [\mnras] {10.1093/mnras/stw878}, \href
  {https://ui.adsabs.harvard.edu/abs/2016MNRAS.459.3963C} {459, 3963}

\bibitem[\protect\citeauthoryear{{Davis}}{{Davis}}{2001}]{2001ApJ...562..575D}
{Davis} J.~E.,  2001, \mn@doi [\apj] {10.1086/323488}, \href
  {https://ui.adsabs.harvard.edu/abs/2001ApJ...562..575D} {562, 575}

\bibitem[\protect\citeauthoryear{{Devereux} \& {Shearer}}{{Devereux} \&
  {Shearer}}{2007}]{2007ApJ...671..118D}
{Devereux} N.,  {Shearer} A.,  2007, \mn@doi [\apj] {10.1086/522292}, \href
  {https://ui.adsabs.harvard.edu/abs/2007ApJ...671..118D} {671, 118}

\bibitem[\protect\citeauthoryear{{Devereux}, {Ford}, {Tsvetanov}  \&
  {Jacoby}}{{Devereux} et~al.}{2003}]{2003AJ....125.1226D}
{Devereux} N.,  {Ford} H.,  {Tsvetanov} Z.,   {Jacoby} G.,  2003, \mn@doi [\aj]
  {10.1086/367595}, \href
  {https://ui.adsabs.harvard.edu/abs/2003AJ....125.1226D} {125, 1226}

\bibitem[\protect\citeauthoryear{{Dexter} et~al.,}{{Dexter}
  et~al.}{2020}]{2020MNRAS.494.4168D}
{Dexter} J.,  et~al., 2020, \mn@doi [\mnras] {10.1093/mnras/staa922}, \href
  {https://ui.adsabs.harvard.edu/abs/2020MNRAS.494.4168D} {494, 4168}

\bibitem[\protect\citeauthoryear{{Dibi}, {Markoff}, {Belmont}, {Malzac},
  {Barri{\`e}re}  \& {Tomsick}}{{Dibi} et~al.}{2014}]{2014MNRAS.441.1005D}
{Dibi} S.,  {Markoff} S.,  {Belmont} R.,  {Malzac} J.,  {Barri{\`e}re} N.~M.,
  {Tomsick} J.~A.,  2014, \mn@doi [\mnras] {10.1093/mnras/stu599}, \href
  {https://ui.adsabs.harvard.edu/abs/2014MNRAS.441.1005D} {441, 1005}

\bibitem[\protect\citeauthoryear{{Dickey} \& {Lockman}}{{Dickey} \&
  {Lockman}}{1990}]{1990ARA&A..28..215D}
{Dickey} J.~M.,  {Lockman} F.~J.,  1990, \mn@doi [\araa]
  {10.1146/annurev.aa.28.090190.001243}, \href
  {https://ui.adsabs.harvard.edu/abs/1990ARA&A..28..215D} {28, 215}

\bibitem[\protect\citeauthoryear{{Dodds-Eden} et~al.,}{{Dodds-Eden}
  et~al.}{2011}]{2011ApJ...728...37D}
{Dodds-Eden} K.,  et~al., 2011, \mn@doi [\apj] {10.1088/0004-637X/728/1/37},
  \href {https://ui.adsabs.harvard.edu/abs/2011ApJ...728...37D} {728, 37}

\bibitem[\protect\citeauthoryear{{Dworetsky}}{{Dworetsky}}{1983}]{1983MNRAS.203..917D}
{Dworetsky} M.~M.,  1983, \mn@doi [\mnras] {10.1093/mnras/203.4.917}, \href
  {https://ui.adsabs.harvard.edu/abs/1983MNRAS.203..917D} {203, 917}

\bibitem[\protect\citeauthoryear{{EHT MWL Science Working Group} et~al.,}{{EHT
  MWL Science Working Group} et~al.}{2021}]{2021ApJ...911L..11E}
{EHT MWL Science Working Group} et~al., 2021, \mn@doi [\apjl]
  {10.3847/2041-8213/abef71}, \href
  {https://ui.adsabs.harvard.edu/abs/2021ApJ...911L..11E} {911, L11}

\bibitem[\protect\citeauthoryear{{Fabian}}{{Fabian}}{2012}]{2012ARA&A..50..455F}
{Fabian} A.~C.,  2012, \mn@doi [\araa] {10.1146/annurev-astro-081811-125521},
  \href {https://ui.adsabs.harvard.edu/abs/2012ARA&A..50..455F} {50, 455}

\bibitem[\protect\citeauthoryear{{Filippenko} \& {Sargent}}{{Filippenko} \&
  {Sargent}}{1988}]{1988ApJ...324..134F}
{Filippenko} A.~V.,  {Sargent} W. L.~W.,  1988, \mn@doi [\apj]
  {10.1086/165886}, \href
  {https://ui.adsabs.harvard.edu/abs/1988ApJ...324..134F} {324, 134}

\bibitem[\protect\citeauthoryear{{Garcia} et~al.,}{{Garcia}
  et~al.}{2010}]{2010ApJ...710..755G}
{Garcia} M.~R.,  et~al., 2010, \mn@doi [\apj] {10.1088/0004-637X/710/1/755},
  \href {https://ui.adsabs.harvard.edu/abs/2010ApJ...710..755G} {710, 755}

\bibitem[\protect\citeauthoryear{{Genzel}, {Sch{\"o}del}, {Ott}, {Eckart},
  {Alexander}, {Lacombe}, {Rouan}  \& {Aschenbach}}{{Genzel}
  et~al.}{2003}]{2003Natur.425..934G}
{Genzel} R.,  {Sch{\"o}del} R.,  {Ott} T.,  {Eckart} A.,  {Alexander} T.,
  {Lacombe} F.,  {Rouan} D.,   {Aschenbach} B.,  2003, \mn@doi [\nat]
  {10.1038/nature02065}, \href
  {https://ui.adsabs.harvard.edu/abs/2003Natur.425..934G} {425, 934}

\bibitem[\protect\citeauthoryear{{George} \& {Fabian}}{{George} \&
  {Fabian}}{1991}]{1991MNRAS.249..352G}
{George} I.~M.,  {Fabian} A.~C.,  1991, \mn@doi [\mnras]
  {10.1093/mnras/249.2.352}, \href
  {https://ui.adsabs.harvard.edu/abs/1991MNRAS.249..352G} {249, 352}

\bibitem[\protect\citeauthoryear{{Ghisellini}}{{Ghisellini}}{2013}]{2013LNP...873.....G}
{Ghisellini} G.,  2013, {Radiative Processes in High Energy Astrophysics}.
 Lecture Notes in Physics Vol. 873, Springer Cham,
  \mn@doi{10.1007/978-3-319-00612-3}

\bibitem[\protect\citeauthoryear{{Haggard} et~al.,}{{Haggard}
  et~al.}{2019}]{2019ApJ...886...96H}
{Haggard} D.,  et~al., 2019, \mn@doi [\apj] {10.3847/1538-4357/ab4a7f}, \href
  {https://ui.adsabs.harvard.edu/abs/2019ApJ...886...96H} {886, 96}

\bibitem[\protect\citeauthoryear{{Heckman}}{{Heckman}}{1980}]{1980A&A....87..152H}
{Heckman} T.~M.,  1980, \aap, \href
  {https://ui.adsabs.harvard.edu/abs/1980A&A....87..152H} {500, 187}

\bibitem[\protect\citeauthoryear{{Heckman} \& {Best}}{{Heckman} \&
  {Best}}{2014}]{2014ARA&A..52..589H}
{Heckman} T.~M.,  {Best} P.~N.,  2014, \mn@doi [\araa]
  {10.1146/annurev-astro-081913-035722}, \href
  {https://ui.adsabs.harvard.edu/abs/2014ARA&A..52..589H} {52, 589}

\bibitem[\protect\citeauthoryear{{Ho}}{{Ho}}{2008}]{2008ARA&A..46..475H}
{Ho} L.~C.,  2008, \mn@doi [\araa] {10.1146/annurev.astro.45.051806.110546},
  \href {https://ui.adsabs.harvard.edu/abs/2008ARA&A..46..475H} {46, 475}

\bibitem[\protect\citeauthoryear{{Ho}, {Filippenko}  \& {Sargent}}{{Ho}
  et~al.}{1996}]{1996ApJ...462..183H}
{Ho} L.~C.,  {Filippenko} A.~V.,   {Sargent} W. L.~W.,  1996, \mn@doi [\apj]
  {10.1086/177140}, \href
  {https://ui.adsabs.harvard.edu/abs/1996ApJ...462..183H} {462, 183}

\bibitem[\protect\citeauthoryear{{Ishisaki} et~al.,}{{Ishisaki}
  et~al.}{1996}]{1996PASJ...48..237I}
{Ishisaki} Y.,  et~al., 1996, \mn@doi [\pasj] {10.1093/pasj/48.2.237}, \href
  {https://ui.adsabs.harvard.edu/abs/1996PASJ...48..237I} {48, 237}

\bibitem[\protect\citeauthoryear{{Iyomoto} \& {Makishima}}{{Iyomoto} \&
  {Makishima}}{2001}]{2001MNRAS.321..767I}
{Iyomoto} N.,  {Makishima} K.,  2001, \mn@doi [\mnras]
  {10.1046/j.1365-8711.2001.04086.x}, \href
  {https://ui.adsabs.harvard.edu/abs/2001MNRAS.321..767I} {321, 767}

\bibitem[\protect\citeauthoryear{{King}, {Miller}, {Bietenholz},
  {G{\"u}ltekin}, {Reynolds}, {Mioduszewski}, {Rupen}  \& {Bartel}}{{King}
  et~al.}{2016}]{2016NatPh..12..772K}
{King} A.~L.,  {Miller} J.~M.,  {Bietenholz} M.,  {G{\"u}ltekin} K.,
  {Reynolds} M.~T.,  {Mioduszewski} A.,  {Rupen} M.,   {Bartel} N.,  2016,
  \mn@doi [Nature Physics] {10.1038/nphys3724}, \href
  {https://ui.adsabs.harvard.edu/abs/2016NatPh..12..772K} {12, 772}

\bibitem[\protect\citeauthoryear{{La Parola}, {Fabbiano}, {Elvis}, {Nicastro},
  {Kim}  \& {Peres}}{{La Parola} et~al.}{2004}]{2004ApJ...601..831L}
{La Parola} V.,  {Fabbiano} G.,  {Elvis} M.,  {Nicastro} F.,  {Kim} D.~W.,
  {Peres} G.,  2004, \mn@doi [\apj] {10.1086/380786}, \href
  {https://ui.adsabs.harvard.edu/abs/2004ApJ...601..831L} {601, 831}

\bibitem[\protect\citeauthoryear{{Li}, {Garcia}, {Forman}, {Jones}, {Kraft},
  {Lal}, {Murray}  \& {Wang}}{{Li} et~al.}{2011}]{2011ApJ...728L..10L}
{Li} Z.,  {Garcia} M.~R.,  {Forman} W.~R.,  {Jones} C.,  {Kraft} R.~P.,  {Lal}
  D.~V.,  {Murray} S.~S.,   {Wang} Q.~D.,  2011, \mn@doi [\apjl]
  {10.1088/2041-8205/728/1/L10}, \href
  {https://ui.adsabs.harvard.edu/abs/2011ApJ...728L..10L} {728, L10}

\bibitem[\protect\citeauthoryear{{Li} et~al.,}{{Li}
  et~al.}{2015}]{2015ApJ...810...19L}
{Li} Y.-P.,  et~al., 2015, \mn@doi [\apj] {10.1088/0004-637X/810/1/19}, \href
  {https://ui.adsabs.harvard.edu/abs/2015ApJ...810...19L} {810, 19}

\bibitem[\protect\citeauthoryear{{Lin}, {Li}  \& {Yuan}}{{Lin}
  et~al.}{2023}]{2023MNRAS.520.1271L}
{Lin} X.,  {Li} Y.-P.,   {Yuan} F.,  2023, \mn@doi [\mnras]
  {10.1093/mnras/stad176}, \href
  {https://ui.adsabs.harvard.edu/abs/2023MNRAS.520.1271L} {520, 1271}

\bibitem[\protect\citeauthoryear{{Ma}, {Roberts}, {Li}  \& {Wang}}{{Ma}
  et~al.}{2019}]{2019MNRAS.483.5614M}
{Ma} R.-Y.,  {Roberts} S.~R.,  {Li} Y.-P.,   {Wang} Q.~D.,  2019, \mn@doi
  [\mnras] {10.1093/mnras/sty3039}, \href
  {https://ui.adsabs.harvard.edu/abs/2019MNRAS.483.5614M} {483, 5614}

\bibitem[\protect\citeauthoryear{{Markoff} et~al.,}{{Markoff}
  et~al.}{2008}]{2008ApJ...681..905M}
{Markoff} S.,  et~al., 2008, \mn@doi [\apj] {10.1086/588718}, \href
  {https://ui.adsabs.harvard.edu/abs/2008ApJ...681..905M} {681, 905}

\bibitem[\protect\citeauthoryear{{McNamara} \& {Nulsen}}{{McNamara} \&
  {Nulsen}}{2012}]{2012NJPh...14e5023M}
{McNamara} B.~R.,  {Nulsen} P.~E.~J.,  2012, \mn@doi [New Journal of Physics]
  {10.1088/1367-2630/14/5/055023}, \href
  {https://ui.adsabs.harvard.edu/abs/2012NJPh...14e5023M} {14, 055023}

\bibitem[\protect\citeauthoryear{{Middleditch}, {Marshall}, {Wang}, {Gotthelf}
  \& {Zhang}}{{Middleditch} et~al.}{2006}]{2006ApJ...652.1531M}
{Middleditch} J.,  {Marshall} F.~E.,  {Wang} Q.~D.,  {Gotthelf} E.~V.,
  {Zhang} W.,  2006, \mn@doi [\apj] {10.1086/508736}, \href
  {https://ui.adsabs.harvard.edu/abs/2006ApJ...652.1531M} {652, 1531}

\bibitem[\protect\citeauthoryear{{Miller}, {Nowak}, {Markoff}, {Rupen}  \&
  {Maitra}}{{Miller} et~al.}{2010}]{2010ApJ...720.1033M}
{Miller} J.~M.,  {Nowak} M.,  {Markoff} S.,  {Rupen} M.~P.,   {Maitra} D.,
  2010, \mn@doi [\apj] {10.1088/0004-637X/720/2/1033}, \href
  {https://ui.adsabs.harvard.edu/abs/2010ApJ...720.1033M} {720, 1033}

\bibitem[\protect\citeauthoryear{{Narayan} \& {Yi}}{{Narayan} \&
  {Yi}}{1994}]{1994ApJ...428L..13N}
{Narayan} R.,  {Yi} I.,  1994, \mn@doi [\apjl] {10.1086/187381}, \href
  {https://ui.adsabs.harvard.edu/abs/1994ApJ...428L..13N} {428, L13}

\bibitem[\protect\citeauthoryear{{Neilsen} et~al.,}{{Neilsen}
  et~al.}{2013}]{2013ApJ...774...42N}
{Neilsen} J.,  et~al., 2013, \mn@doi [\apj] {10.1088/0004-637X/774/1/42}, \href
  {https://ui.adsabs.harvard.edu/abs/2013ApJ...774...42N} {774, 42}

\bibitem[\protect\citeauthoryear{{Nemmen}, {Storchi-Bergmann}  \&
  {Eracleous}}{{Nemmen} et~al.}{2014}]{2014MNRAS.438.2804N}
{Nemmen} R.~S.,  {Storchi-Bergmann} T.,   {Eracleous} M.,  2014, \mn@doi
  [\mnras] {10.1093/mnras/stt2388}, \href
  {https://ui.adsabs.harvard.edu/abs/2014MNRAS.438.2804N} {438, 2804}

\bibitem[\protect\citeauthoryear{{Nied{\'z}wiecki}, {Xie}  \&
  {Zdziarski}}{{Nied{\'z}wiecki} et~al.}{2012}]{2012MNRAS.420.1195N}
{Nied{\'z}wiecki} A.,  {Xie} F.-G.,   {Zdziarski} A.~A.,  2012, \mn@doi
  [\mnras] {10.1111/j.1365-2966.2011.20106.x}, \href
  {https://ui.adsabs.harvard.edu/abs/2012MNRAS.420.1195N} {420, 1195}

\bibitem[\protect\citeauthoryear{{Page}, {Breeveld}, {Soria}, {Wu}, {Brand
  uardi-Raymont}, {Mason}, {Starling}  \& {Zane}}{{Page}
  et~al.}{2003}]{2003A&A...400..145P}
{Page} M.~J.,  {Breeveld} A.~A.,  {Soria} R.,  {Wu} K.,  {Brand uardi-Raymont}
  G.,  {Mason} K.~O.,  {Starling} R.~L.~C.,   {Zane} S.,  2003, \mn@doi [\aap]
  {10.1051/0004-6361:20021896}, \href
  {https://ui.adsabs.harvard.edu/abs/2003A&A...400..145P} {400, 145}

\bibitem[\protect\citeauthoryear{{Page}, {Soria}, {Zane}, {Wu}  \&
  {Starling}}{{Page} et~al.}{2004}]{2004A&A...422...77P}
{Page} M.~J.,  {Soria} R.,  {Zane} S.,  {Wu} K.,   {Starling} R.~L.~C.,  2004,
  \mn@doi [\aap] {10.1051/0004-6361:20034451}, \href
  {https://ui.adsabs.harvard.edu/abs/2004A&A...422...77P} {422, 77}

\bibitem[\protect\citeauthoryear{{Peimbert} \& {Torres-Peimbert}}{{Peimbert} \&
  {Torres-Peimbert}}{1981}]{1981ApJ...245..845P}
{Peimbert} M.,  {Torres-Peimbert} S.,  1981, \mn@doi [\apj] {10.1086/158860},
  \href {https://ui.adsabs.harvard.edu/abs/1981ApJ...245..845P} {245, 845}

\bibitem[\protect\citeauthoryear{{Pellegrini}, {Cappi}, {Bassani}, {Malaguti},
  {Palumbo}  \& {Persic}}{{Pellegrini} et~al.}{2000}]{2000A&A...353..447P}
{Pellegrini} S.,  {Cappi} M.,  {Bassani} L.,  {Malaguti} G.,  {Palumbo}
  G.~G.~C.,   {Persic} M.,  2000, \aap, \href
  {https://ui.adsabs.harvard.edu/abs/2000A&A...353..447P} {353, 447}

\bibitem[\protect\citeauthoryear{{Ponti} et~al.,}{{Ponti}
  et~al.}{2015}]{2015MNRAS.453..172P}
{Ponti} G.,  et~al., 2015, \mn@doi [\mnras] {10.1093/mnras/stv1331}, \href
  {https://ui.adsabs.harvard.edu/abs/2015MNRAS.453..172P} {453, 172}

\bibitem[\protect\citeauthoryear{{Ponti} et~al.,}{{Ponti}
  et~al.}{2017}]{2017MNRAS.468.2447P}
{Ponti} G.,  et~al., 2017, \mn@doi [\mnras] {10.1093/mnras/stx596}, \href
  {https://ui.adsabs.harvard.edu/abs/2017MNRAS.468.2447P} {468, 2447}

\bibitem[\protect\citeauthoryear{{Prieto}, {Fern{\'a}ndez-Ontiveros},
  {Markoff}, {Espada}  \& {Gonz{\'a}lez-Mart{\'\i}n}}{{Prieto}
  et~al.}{2016}]{2016MNRAS.457.3801P}
{Prieto} M.~A.,  {Fern{\'a}ndez-Ontiveros} J.~A.,  {Markoff} S.,  {Espada} D.,
   {Gonz{\'a}lez-Mart{\'\i}n} O.,  2016, \mn@doi [\mnras]
  {10.1093/mnras/stw166}, \href
  {https://ui.adsabs.harvard.edu/abs/2016MNRAS.457.3801P} {457, 3801}

\bibitem[\protect\citeauthoryear{{Quataert}, {Di Matteo}, {Narayan}  \&
  {Ho}}{{Quataert} et~al.}{1999}]{1999ApJ...525L..89Q}
{Quataert} E.,  {Di Matteo} T.,  {Narayan} R.,   {Ho} L.~C.,  1999, \mn@doi
  [\apjl] {10.1086/312353}, \href
  {https://ui.adsabs.harvard.edu/abs/1999ApJ...525L..89Q} {525, L89}

\bibitem[\protect\citeauthoryear{{Roberts}, {Jiang}, {Wang}  \&
  {Ostriker}}{{Roberts} et~al.}{2017}]{2017MNRAS.466.1477R}
{Roberts} S.~R.,  {Jiang} Y.-F.,  {Wang} Q.~D.,   {Ostriker} J.~P.,  2017,
  \mn@doi [\mnras] {10.1093/mnras/stw2995}, \href
  {https://ui.adsabs.harvard.edu/abs/2017MNRAS.466.1477R} {466, 1477}

\bibitem[\protect\citeauthoryear{{Sell}, {Pooley}, {Zezas}, {Heinz}, {Homan}
  \& {Lewin}}{{Sell} et~al.}{2011}]{2011ApJ...735...26S}
{Sell} P.~H.,  {Pooley} D.,  {Zezas} A.,  {Heinz} S.,  {Homan} J.,   {Lewin}
  W.~H.~G.,  2011, \mn@doi [\apj] {10.1088/0004-637X/735/1/26}, \href
  {https://ui.adsabs.harvard.edu/abs/2011ApJ...735...26S} {735, 26}

\bibitem[\protect\citeauthoryear{{Shakura} \& {Sunyaev}}{{Shakura} \&
  {Sunyaev}}{1973}]{1973A&A....24..337S}
{Shakura} N.~I.,  {Sunyaev} R.~A.,  1973, \aap, \href
  {https://ui.adsabs.harvard.edu/abs/1973A&A....24..337S} {500, 33}

\bibitem[\protect\citeauthoryear{{Shi}, {Li}, {Yuan}  \& {Zhu}}{{Shi}
  et~al.}{2021}]{2021NatAs...5..928S}
{Shi} F.,  {Li} Z.,  {Yuan} F.,   {Zhu} B.,  2021, \mn@doi [Nature Astronomy]
  {10.1038/s41550-021-01394-0}, \href
  {https://ui.adsabs.harvard.edu/abs/2021NatAs...5..928S} {5, 928}

\bibitem[\protect\citeauthoryear{{Swartz}, {Ghosh}, {McCollough}, {Pannuti},
  {Tennant}  \& {Wu}}{{Swartz} et~al.}{2003}]{2003ApJS..144..213S}
{Swartz} D.~A.,  {Ghosh} K.~K.,  {McCollough} M.~L.,  {Pannuti} T.~G.,
  {Tennant} A.~F.,   {Wu} K.,  2003, \mn@doi [\apjs] {10.1086/345084}, \href
  {https://ui.adsabs.harvard.edu/abs/2003ApJS..144..213S} {144, 213}

\bibitem[\protect\citeauthoryear{{Tully}, {Courtois}  \& {Sorce}}{{Tully}
  et~al.}{2016}]{2016AJ....152...50T}
{Tully} R.~B.,  {Courtois} H.~M.,   {Sorce} J.~G.,  2016, \mn@doi [\aj]
  {10.3847/0004-6256/152/2/50}, \href
  {https://ui.adsabs.harvard.edu/abs/2016AJ....152...50T} {152, 50}

\bibitem[\protect\citeauthoryear{{Verner}, {Ferland}, {Korista}  \&
  {Yakovlev}}{{Verner} et~al.}{1996}]{1996ApJ...465..487V}
{Verner} D.~A.,  {Ferland} G.~J.,  {Korista} K.~T.,   {Yakovlev} D.~G.,  1996,
  \mn@doi [\apj] {10.1086/177435}, \href
  {https://ui.adsabs.harvard.edu/abs/1996ApJ...465..487V} {465, 487}

\bibitem[\protect\citeauthoryear{{Wang} et~al.,}{{Wang}
  et~al.}{2013}]{2013Sci...341..981W}
{Wang} Q.~D.,  et~al., 2013, \mn@doi [Science] {10.1126/science.1240755}, \href
  {https://ui.adsabs.harvard.edu/abs/2013Sci...341..981W} {341, 981}

\bibitem[\protect\citeauthoryear{{Wilms}, {Allen}  \& {McCray}}{{Wilms}
  et~al.}{2000}]{2000ApJ...542..914W}
{Wilms} J.,  {Allen} A.,   {McCray} R.,  2000, \mn@doi [\apj] {10.1086/317016},
  \href {https://ui.adsabs.harvard.edu/abs/2000ApJ...542..914W} {542, 914}

\bibitem[\protect\citeauthoryear{{Witzel} et~al.,}{{Witzel}
  et~al.}{2018}]{2018ApJ...863...15W}
{Witzel} G.,  et~al., 2018, \mn@doi [\apj] {10.3847/1538-4357/aace62}, \href
  {https://ui.adsabs.harvard.edu/abs/2018ApJ...863...15W} {863, 15}

\bibitem[\protect\citeauthoryear{{Xie} \& {Yuan}}{{Xie} \&
  {Yuan}}{2012}]{2012MNRAS.427.1580X}
{Xie} F.-G.,  {Yuan} F.,  2012, \mn@doi [\mnras]
  {10.1111/j.1365-2966.2012.22030.x}, \href
  {https://ui.adsabs.harvard.edu/abs/2012MNRAS.427.1580X} {427, 1580}

\bibitem[\protect\citeauthoryear{{Xie} \& {Yuan}}{{Xie} \&
  {Yuan}}{2017}]{2017ApJ...836..104X}
{Xie} F.-G.,  {Yuan} F.,  2017, \mn@doi [\apj] {10.3847/1538-4357/aa5b90},
  \href {https://ui.adsabs.harvard.edu/abs/2017ApJ...836..104X} {836, 104}

\bibitem[\protect\citeauthoryear{{Xie}, {Narayan}  \& {Yuan}}{{Xie}
  et~al.}{2023}]{2023ApJ...942...20X}
{Xie} F.-G.,  {Narayan} R.,   {Yuan} F.,  2023, \mn@doi [\apj]
  {10.3847/1538-4357/aca534}, \href
  {https://ui.adsabs.harvard.edu/abs/2023ApJ...942...20X} {942, 20}

\bibitem[\protect\citeauthoryear{{Xu} \& {Cao}}{{Xu} \&
  {Cao}}{2009}]{2009RAA.....9..401X}
{Xu} Y.-D.,  {Cao} X.-W.,  2009, \mn@doi [Research in Astronomy and
  Astrophysics] {10.1088/1674-4527/9/4/003}, \href
  {https://ui.adsabs.harvard.edu/abs/2009RAA.....9..401X} {9, 401}

\bibitem[\protect\citeauthoryear{{Yang}, {Xie}, {Yuan}, {Zdziarski},
  {Gierli{\'n}ski}, {Ho}  \& {Yu}}{{Yang} et~al.}{2015}]{2015MNRAS.447.1692Y}
{Yang} Q.-X.,  {Xie} F.-G.,  {Yuan} F.,  {Zdziarski} A.~A.,  {Gierli{\'n}ski}
  M.,  {Ho} L.~C.,   {Yu} Z.,  2015, \mn@doi [\mnras] {10.1093/mnras/stu2571},
  \href {https://ui.adsabs.harvard.edu/abs/2015MNRAS.447.1692Y} {447, 1692}

\bibitem[\protect\citeauthoryear{{Young}, {Nowak}, {Markoff}, {Marshall}  \&
  {Canizares}}{{Young} et~al.}{2007}]{2007ApJ...669..830Y}
{Young} A.~J.,  {Nowak} M.~A.,  {Markoff} S.,  {Marshall} H.~L.,   {Canizares}
  C.~R.,  2007, \mn@doi [\apj] {10.1086/521778}, \href
  {https://ui.adsabs.harvard.edu/abs/2007ApJ...669..830Y} {669, 830}

\bibitem[\protect\citeauthoryear{{Young}, {McHardy}, {Emmanoulopoulos}  \&
  {Connolly}}{{Young} et~al.}{2018}]{2018MNRAS.476.5698Y}
{Young} A.~J.,  {McHardy} I.,  {Emmanoulopoulos} D.,   {Connolly} S.,  2018,
  \mn@doi [\mnras] {10.1093/mnras/sty509}, \href
  {https://ui.adsabs.harvard.edu/abs/2018MNRAS.476.5698Y} {476, 5698}

\bibitem[\protect\citeauthoryear{{Yuan} \& {Narayan}}{{Yuan} \&
  {Narayan}}{2004}]{2004ApJ...612..724Y}
{Yuan} F.,  {Narayan} R.,  2004, \mn@doi [\apj] {10.1086/422802}, \href
  {https://ui.adsabs.harvard.edu/abs/2004ApJ...612..724Y} {612, 724}

\bibitem[\protect\citeauthoryear{{Yuan} \& {Narayan}}{{Yuan} \&
  {Narayan}}{2014}]{2014ARA&A..52..529Y}
{Yuan} F.,  {Narayan} R.,  2014, \mn@doi [\araa]
  {10.1146/annurev-astro-082812-141003}, \href
  {https://ui.adsabs.harvard.edu/abs/2014ARA&A..52..529Y} {52, 529}

\bibitem[\protect\citeauthoryear{{Yuan} \& {Wang}}{{Yuan} \&
  {Wang}}{2016}]{2016MNRAS.456.1438Y}
{Yuan} Q.,  {Wang} Q.~D.,  2016, \mn@doi [\mnras] {10.1093/mnras/stv2778},
  \href {https://ui.adsabs.harvard.edu/abs/2016MNRAS.456.1438Y} {456, 1438}

\bibitem[\protect\citeauthoryear{{Yuan}, {Lin}, {Wu}  \& {Ho}}{{Yuan}
  et~al.}{2009}]{2009MNRAS.395.2183Y}
{Yuan} F.,  {Lin} J.,  {Wu} K.,   {Ho} L.~C.,  2009, \mn@doi [\mnras]
  {10.1111/j.1365-2966.2009.14673.x}, \href
  {https://ui.adsabs.harvard.edu/abs/2009MNRAS.395.2183Y} {395, 2183}

\bibitem[\protect\citeauthoryear{{Yuan}, {Bu}  \& {Wu}}{{Yuan}
  et~al.}{2012}]{2012ApJ...761..130Y}
{Yuan} F.,  {Bu} D.,   {Wu} M.,  2012, \mn@doi [\apj]
  {10.1088/0004-637X/761/2/130}, \href
  {https://ui.adsabs.harvard.edu/abs/2012ApJ...761..130Y} {761, 130}

\bibitem[\protect\citeauthoryear{{Yuan}, {Gan}, {Narayan}, {Sadowski}, {Bu}  \&
  {Bai}}{{Yuan} et~al.}{2015}]{2015ApJ...804..101Y}
{Yuan} F.,  {Gan} Z.,  {Narayan} R.,  {Sadowski} A.,  {Bu} D.,   {Bai} X.-N.,
  2015, \mn@doi [\apj] {10.1088/0004-637X/804/2/101}, \href
  {https://ui.adsabs.harvard.edu/abs/2015ApJ...804..101Y} {804, 101}

\bibitem[\protect\citeauthoryear{{Yuan}, {Wang}, {Liu}  \& {Wu}}{{Yuan}
  et~al.}{2018}]{2018MNRAS.473..306Y}
{Yuan} Q.,  {Wang} Q.~D.,  {Liu} S.,   {Wu} K.,  2018, \mn@doi [\mnras]
  {10.1093/mnras/stx2408}, \href
  {https://ui.adsabs.harvard.edu/abs/2018MNRAS.473..306Y} {473, 306}

\bibitem[\protect\citeauthoryear{{Yusef-Zadeh}, {Roberts}, {Wardle}, {Heinke}
  \& {Bower}}{{Yusef-Zadeh} et~al.}{2006}]{2006ApJ...650..189Y}
{Yusef-Zadeh} F.,  {Roberts} D.,  {Wardle} M.,  {Heinke} C.~O.,   {Bower}
  G.~C.,  2006, \mn@doi [\apj] {10.1086/506375}, \href
  {https://ui.adsabs.harvard.edu/abs/2006ApJ...650..189Y} {650, 189}

\bibitem[\protect\citeauthoryear{{Yusef-Zadeh} et~al.,}{{Yusef-Zadeh}
  et~al.}{2009}]{2009ApJ...702..178Y}
{Yusef-Zadeh} F.,  et~al., 2009, \mn@doi [\apj] {10.1088/0004-637X/702/1/178},
  \href {https://ui.adsabs.harvard.edu/abs/2009ApJ...702..178Y} {702, 178}

\bibitem[\protect\citeauthoryear{{Yusef-Zadeh}, {Wardle}, {Kunneriath},
  {Royster}, {Wootten}  \& {Roberts}}{{Yusef-Zadeh}
  et~al.}{2017}]{2017ApJ...850L..30Y}
{Yusef-Zadeh} F.,  {Wardle} M.,  {Kunneriath} D.,  {Royster} M.,  {Wootten} A.,
    {Roberts} D.~A.,  2017, \mn@doi [\apjl] {10.3847/2041-8213/aa96a2}, \href
  {https://ui.adsabs.harvard.edu/abs/2017ApJ...850L..30Y} {850, L30}

\bibitem[\protect\citeauthoryear{{Zhang} et~al.,}{{Zhang}
  et~al.}{2017}]{2017ApJ...843...96Z}
{Zhang} S.,  et~al., 2017, \mn@doi [\apj] {10.3847/1538-4357/aa74e8}, \href
  {https://ui.adsabs.harvard.edu/abs/2017ApJ...843...96Z} {843, 96}

\bibitem[\protect\citeauthoryear{{{\v{C}}emelji{\'c}}, {Yang}, {Yuan}  \&
  {Shang}}{{{\v{C}}emelji{\'c}} et~al.}{2022}]{2022ApJ...933...55C}
{{\v{C}}emelji{\'c}} M.,  {Yang} H.,  {Yuan} F.,   {Shang} H.,  2022, \mn@doi
  [\apj] {10.3847/1538-4357/ac70cc}, \href
  {https://ui.adsabs.harvard.edu/abs/2022ApJ...933...55C} {933, 55}

\makeatother
\end{thebibliography}

  
  
  
  %
  %
  
  \bsp  
  \label{lastpage}
\end{document}